\documentclass[12pt,amssymb,qsymbols]{article}

\pdfoutput=1
\usepackage{tabularx}
\usepackage[pdftex]{graphicx}
\usepackage{amsmath}
\usepackage{amssymb}
\usepackage{longtable}
\usepackage{verbatim}
\usepackage{hyperref}
\usepackage{color}

\makeatletter

\setlongtables

\newcommand{\bea}{\begin{eqnarray}}
\newcommand{\eea}{\end{eqnarray}}
\newcommand{\beq}{\begin{equation}}
\newcommand{\eeq}{\end{equation}}
\newcommand{\ec}{\end{center}}
\newcommand{\bc}{\begin{center}}

\newcommand{\pdir}{p\kern -5.2pt\raise 0.2ex\hbox {/}}

\newcommand{\vdir}{v\kern -5.75pt\raise 0.15ex\hbox {/}}
\newcommand{\kdir}{k\kern -5.75pt\raise 0.15ex\hbox {/}}
\newcommand{\epsdir}{\epsilon\kern -5.0pt\raise 0.15ex\hbox {/}}
\newcommand{\bvdir}{\bar{v}\kern -5.75pt\raise 0.15ex\hbox {/}}
\newcommand{\Ddir}{D\kern -7.75pt\raise 0.20ex\hbox {/}}
\newcommand{\Adir}{A\kern -7.75pt\raise 0.20ex\hbox {/}}
\newcommand{\ldir}{l\kern -5.0pt\raise 0.2ex\hbox{/}}
\newcommand{\varepsdir}{\varepsilon\kern -5.5pt\raise 0.15ex\hbox{/}}

\def \eff{{\text{eff}}}

\newcommand{\nn}{\nonumber}
\makeatother

\newcommand{\ord}{\mathcal{O}}

\newcommand{\eq}[1]{\begin{equation} #1 \end{equation}}
\newcommand{\eqa}[1]{\begin{eqnarray} #1 \end{eqnarray}}
\newcommand{\C}[1]{{\cal C}_{#1}}

\newcommand{\av}[1]{\langle #1 \rangle}

\definecolor{orange}{rgb}{1.0,0.5,0}

\definecolor{nicered}{rgb}{0.7,0.1,0.1}
\definecolor{nicegreen}{rgb}{0.1,0.5,0.1}
\hypersetup{colorlinks,citecolor= nicegreen,linkcolor= nicered}

\def\draftdate{\relax}
\def\mda{\relax}
\def\mua{\relax}
\def\mla{\relax}
\def\draft{
\def\thtystars{******************************}
\def\sixtystars{\thtystars\thtystars}
\typeout{}
\typeout{\sixtystars**}
\typeout{* Draft mode!
         For final version remove \protect\draft\space in source file *}
\typeout{\sixtystars**}
\typeout{}
\def\draftdate{\today}
\def\mua{\marginpar[\boldmath\hfil$\uparrow$]%
                   {\boldmath$\uparrow$\hfil}%
                    \typeout{marginpar: $\uparrow$}\ignorespaces}
\def\mda{\marginpar[\boldmath\hfil$\downarrow$]%
                   {\boldmath$\downarrow$\hfil}%
                    \typeout{marginpar: $\downarrow$}\ignorespaces}
\def\mla{\marginpar[\boldmath\hfil$\rightarrow$]%
                   {\boldmath$\leftarrow $\hfil}%
                    \typeout{marginpar: $\leftrightarrow$}\ignorespaces}
\def\Mua{\marginpar[\boldmath\hfil$\Uparrow$]%
                   {\boldmath$\Uparrow$\hfil}%
                    \typeout{marginpar: $\uparrow$}\ignorespaces}
\def\Mda{\marginpar[\boldmath\hfil$\Downarrow$]%
                   {\boldmath$\Downarrow$\hfil}%
                    \typeout{marginpar: $\downarrow$}\ignorespaces}
\def\Mla{\marginpar[\boldmath\hfil$\Rightarrow$]%
                   {\boldmath$\Leftarrow $\hfil}%
                    \typeout{marginpar: $\leftrightarrow$}\ignorespaces}
\overfullrule 5pt
\oddsidemargin -15mm
\marginparwidth 29mm
}

\begin{document}

\begin{flushright}
{\small
LPT-ORSAY/14-63\\
UAB-FT-757\\
QFET-2014-13\\
SI-HEP-2014-19 
}
\end{flushright}
$\ $
\vspace{0cm}

\begin{center}
\Large\bf
On the impact of power corrections\\ in the prediction of $B \to K^* \mu^+\mu^-$ observables
\end{center}

\vspace{0mm}
\begin{center}
{\sc S\'ebastien Descotes-Genon}\\[2mm]
{\em\small
Laboratoire de Physique Th\'eorique, CNRS/Univ. Paris-Sud 11 (UMR 8627)\\ 91405 Orsay Cedex, France
}
\end{center}

\begin{center}
{\sc  Lars Hofer, Joaquim Matias}\\[2mm]
{\em \small
Universitat Aut\`onoma de Barcelona, 08193 Bellaterra, Barcelona
}
\end{center}

\begin{center}
{\sc  Javier Virto}\\[2mm]
{\em \small Theoretische Physik 1, Naturwissenschaftlich-Technische Fakult\"at,\\
Universit\"at Siegen, 57068 Siegen, Germany
}
\end{center}

\vspace{3mm}

\begin{abstract}
The recent LHCb angular analysis of the exclusive decay $B\to K^* \mu^+ \mu^-$ has indicated significant deviations from the
Standard Model expectations. Accurate predictions can be achieved at large $K^*$-meson recoil for an optimised set of observables designed to have no sensitivity to hadronic input in the heavy-quark limit at leading order in  $\alpha_s$. However, hadronic uncertainties reappear through non-perturbative $\Lambda_{\rm QCD}/m_b$ power corrections, which must be assessed precisely. In the framework of QCD factorisation we present a systematic method to include 
factorisable power corrections and point out that their impact on
angular observables depends on the scheme chosen to define the soft form factors. 
Associated uncertainties are found to be under control, contrary to earlier claims
in the literature. We also discuss the impact of possible non-factorisable power corrections, including an estimate of charm-loop effects. 
We provide results for angular observables at large recoil  for two different sets of inputs for the form factors,
spelling out the different sources of theoretical uncertainties.
Finally, we comment on a recent proposal to explain the anomaly in $B\to K^* \mu^+ \mu^-$ observables through charm-resonance
effects, and we propose strategies to test this proposal 
identifying observables and kinematic regions where either the charm-loop model can be disentangled from New Physics effects or the two options leave different imprints.
\end{abstract}


\newpage

\section{Introduction}

Since many years, radiative transitions $b\to s\gamma^{(*)}$ have been considered as very powerful probes of physics beyond the Standard Model (SM). These Flavour-Changing Neutral Currents (FCNC) are only mediated by loops in the Standard Model and thus exhibit a quantum
sensitivity to New Physics (NP). Recently the exclusive decay $B\to K^*\mu^+\mu^-$ has come to prominence, as the latest LHCb angular
analysis~\cite{Aaij:2013iag,Aaij:2013qta} suggests significant deviations from the Standard Model, most notably in the observables
$P_5'$~\cite{DescotesGenon:2012zf} and $P_2$~\cite{Becirevic:2011bp,Matias:2012xw} in the region of large hadronic recoil.
Within the model-independent effective Hamiltonian approach (summarised for instance in ref.~\cite{Altmannshofer:2008dz}), it has been shown in ref.~\cite{Descotes-Genon:2013wba}
that such deviations can be easily accommodated in the presence of short-distance NP contributions to the semileptonic operator
$O_9$, reducing the size of the Wilson coefficient $C_9$ by 25\% with respect to the SM prediction.

Several later studies have reached similar conclusions from $B\to K^*\mu^+\mu^-$ data, using different
observables~\cite{Altmannshofer:2013foa} and/or statistical methods~\cite{Beaujean:2013soa}, with possible interpretations in
terms of $Z'$ models\footnote{
Such a model would also have an impact on purely hadronic $B$ and $B_s$ decays. It could explain
the tension in $\Delta A_{\rm CP}$ in $B\to\pi K$ decays and lead to a large enhancement above the SM expectations of the
branching ratios of the purely isospin-violating decays $B_s\to\phi\pi$ and $B_s\to \phi\rho$~\cite{Hofer:2010ee,Hofer:2012vc}.
The anomaly has also been addressed in the context of other NP models
\cite{Altmannshofer:2013foa,Altmannshofer:2014cfa,Mahmoudi:2014mja,Biancofiore:2014wpa,Buras:2014yna,Datta:2013kja}
and within MFV~\cite{Hurth:2013ssa}. 
}~\cite{Descotes-Genon:2013wba,Altmannshofer:2013foa,Gauld:2013qba,Gauld:2013qja,Buras:2013dea,Buras:2013qja}.
Interestingly, a recent lattice study of $B\to K^*\mu^+\mu^-$ and $B_s\to\phi\mu^+\mu^-$ form
factors~\cite{Horgan:2013hoa,Horgan:2013pva} confirms the same trend using different observables (branching ratios rather than angular observables) and  in a different kinematic regime (low rather than large hadronic recoil).
The need for NP contributions to other operators, and in particular to the chirality-flipped semileptonic
operator $O_{9}^{\prime}$, is currently debated depending on the subset of observables and bins
chosen and the input for hadronic form factors~\cite{Descotes-Genon:2013wba,Altmannshofer:2013foa,Beaujean:2013soa,Horgan:2013pva,proc}. Another issue currently debated is the exact role of
long-distance $c\bar{c}$ loops, for which only partial estimates exist~\cite{Khodjamirian:2010vf}, pushing in the opposite 
direction to LHCb data, i.e. making the anomaly more severe~\cite{Descotes-Genon:2013wba}. 
A comparison of BES data on $\sigma(e^+e^-\to$ hadrons)~\cite{Bai:2001ct} and the
$B^+\to K^+\mu^+\mu^-$ dimuon spectrum~\cite{Aaij:2014tfa} suggests that non-factorisable corrections above the $c\bar c$ threshold are
very large. Dispersive approaches have been used to exploit this information and to estimate the effects in
$B\to K^*\mu^+\mu^-$~\cite{zwicky}, but it remains unclear how reliable these methods are in reconstructing the needed
real and imaginary parts of the $c\bar{c}$ loop function (including all long-distance effects) at low recoil, as well as its extrapolation to the large-recoil region (see Appendix for futher discussion).

A usual problem in quark flavour physics is the precise estimation of hadronic uncertainties, necessary for a correct
comparison between theory and data. Reliable constraints on short-distance Wilson coefficients depend on that
premise, as well as  the statistical assessment of deviations potentially related to NP. This can only be achieved if long-distance effects,
encoded for example in hadronic form factors, are sufficiently under control. Although calculations with different
non-perturbative methods are available, e.g., light-cone sum rules (LCSR) at large recoil and lattice QCD at low recoil, they have
not yet reached an accuracy matching the experimental measurements. In addition, as decay amplitudes combine
different form factors (in their canonical definition from $B\to K^*$ vector and tensor matrix elements), the absence of
proper assessment of correlations among the different form factors can lead to a significant enhancement in the uncertainties
of SM predictions of $B\to K^*$ decay amplitudes, and thus of the decay rate and its angular coefficients.

A fruitful approach to this problem has consisted in identifying observables built as suitable combinations of
angular coefficients, where hadronic uncertainties cancel to a large extent (so-called form-factor independent or optimised
observables). A guiding principle has
been the use of effective theories (QCD factorisation/Soft Collinear Effective Theory at large recoil,
Heavy Quark Effective Theory at low recoil)~\cite{Beneke:2000wa, Beneke:2001at, Grinstein:2004vb}, allowing one to separate
hard physics (occurring at scales around $m_b$) and soft physics (around and below $\Lambda_{\rm QCD}$) through an expansion of the form
factors in $\Lambda/m_b$. The $B\to K^*$ decay amplitudes and related angular coefficients can be analysed through similar expansions, based on the factorisation of the seven QCD form
factors in terms of only two soft form factors $\xi_\perp$ and $\xi_{||}$~\cite{Beneke:2000wa,Charles:1998dr,Bauer:2000yr}.
In this context, form-factor-independent observables are defined as observables where the soft form factors cancel at leading
order of the effective theory for the kinematic regime of interest (low or large $K^*$ recoil). This has led for instance to the
transverse asymmetries $A^{(i)}_T$~\cite{Becirevic:2011bp,Kruger:2005ep,Egede:2008uy,Egede:2010zc} and later to the observables
$P_i^{(\prime)}$~\cite{DescotesGenon:2012zf,Matias:2012xw} at large $K^*$ recoil\footnote{
Similar observables can be built at low recoil~\cite{Bobeth:2010wg,Bobeth:2011gi,Descotes-Genon:2013vna},
but the structure of the form-factor relations is different due to the different effective theory holding in this kinematic regime.
}.
It was shown that a clever choice of observables could drastically reduce the sensitivity to hadronic inputs and enhance the
sensitivity to New Physics~\cite{Descotes-Genon:2013vna}.

Beyond leading order the above-mentioned decomposition of the seven QCD form
factors in terms of two soft form factors receives $\alpha_s$ corrections (coming from hard-gluon exchanges) and $1/m_b$ power corrections (due to soft-gluon exchanges)~\cite{Beneke:2000wa,Bauer:2000yr}. In the QCD factorisation analysis of $B\to K^*\mu^+\mu^-$ at large recoil~\cite{Beneke:2001at}, where amplitudes are expressed in terms of Wilson coefficients and soft form factors, these corrections to the relation between QCD and soft form factors manifest themselves as so-called \emph{factorisable corrections}. 
The QCD factorisation analysis of the $B\to K^*\mu^+\mu^-$ amplitudes leads to further $\alpha_s$ and $1/m_b$ corrections called \emph{non-factorisable corrections}, which are not related to form factors, for instance those coming from four-quark operators that can be inserted  in the $B\to K^*\mu^+\mu^-$ decay (forming a $c\bar{c}$ pair decaying into a dimuon pair).

A first approach to predict $B\to K^*\mu^+\mu^-$  observables in the large-recoil region is naive factorisation, using the seven full (QCD) form factors for the three operators $O_{7}$ (electromagnetic), $O_9$ and $O_{10}$ (semileptonic), but neglecting
effects from four-quark operators beyond their high-energy contribution accounted for by the effective Wilson coefficients
$C_{7,9}^{\rm eff}$ (see refs.~\cite{Altmannshofer:2008dz,DescotesGenon:2011yn} for the definition of the operators and Wilson coefficients).  If the form factors are computed fully non-perturbatively, predictions obtained in this way include \emph{factorisable}  $\alpha_s$ and $1/m_b$ power corrections to all orders.
The method can be extended
beyond naive factorisation by adding perturbative $\mathcal{O}(\alpha_s)$ corrections to the contribution
from four-quark operators within the framework of QCD factorisation~\cite{Beneke:2001at}, as in ref.~\cite{Altmannshofer:2008dz}.
These corrections rely on the factorisation of matrix elements of hadronic operators at the leading power in
a $(\Lambda/E_{K^*},\Lambda/m_b)$ expansion. Contributions from four-quark operators at subleading power are not known and must be estimated.
We will refer to these contributions as \emph{non-factorisable power corrections}.
As a downside of this approach, the form factor dependence does not cancel analytically in optimised observables, and
in order to obtain accurate predictions it is crucial to know precisely the correlations among the uncertainties of the different
form factors. In practice, however, LCSR results are usually presented without specifying the correlations among the various
form factors. Moreover, while in principle parametric correlations originating from the hadronic inputs can be traced back easily,
more sophisticated intrinsic  correlations (e.g., the dependence on the Borel parameter) are hard to pin down.

A second approach consists in factorising the QCD form factors using effective field theory
methods~\cite{Beneke:2000wa,Charles:1998dr,Bauer:2000yr}. In this approach, correlations among the
seven QCD form factors are to a large extent accounted for by their expression in terms of the two soft form factors
$\xi_\perp$ and $\xi_{||}$. At leading order in $\alpha_s$ (and at leading $1/m_b$ power) this leads to an analytic cancellation
of form factors within optimised observables.
Hence this method enables one to obtain
precise predictions even in the absence of a precise knowledge on the correlations among the form factors. 
The dependence on form factors obviously reappears through (factorisable and non-factorisable) corrections to the leading-order results, either via
(perturbative and calculable) $\mathcal{O}(\alpha_s)$- or (non-perturbative) $\mathcal{O}(\Lambda/m_b)$-corrections
 to the factorisation formula for QCD form factors. 
The fact that the dominant errors from form factors are suppressed in form-factor independent observables by one power
of $\alpha_s$ or $\Lambda/m_b$ makes these 
observables quite sensitive to subleading $\Lambda/m_b$ power corrections (either factorisable or non-factorisable). 

In order to determine the significance of the deviations in $B\to K^*\mu^+\mu^-$ with respect to the Standard Model,
it appears thus essential to estimate the size of the $\Lambda/m_b$ power corrections. While an estimate of factorisable
corrections is needed to get a reliable prediction from the second method, non-factorisable corrections have to be considered
in both approaches. Non-factorisable power corrections cannot be computed from first principles, but factorisable ones
can be extracted from QCD form factors by separating the contribution from soft form factors.
This issue was discussed recently in ref.~\cite{Jager:2012uw}, 
suggesting that factorisable power corrections estimated in this way
would imply substantial hadronic uncertainties on $B\to K^*\mu^+\mu^-$
observables, much larger than what was found in other works. The present paper aims at reassessing these claims, showing that
these large uncertainties are largely due to peculiar choices in the analysis method used in ref.~\cite{Jager:2012uw} and are not a consequence of the theoretical information currently available on $B\to K^*$ form factors.

The paper is organized as follows. We begin in Section~\ref{sec:LO} by describing the decomposition of QCD form factors
in terms of soft form factors, including perturbative and power corrections, and discussing the role of the renormalisation scheme.
In Section~\ref{sec:FacPow} we describe our approach to factorisable power corrections, leading to our estimates for power
correction parameters and their uncertainties. We then discuss the impact of
these power corrections in the binned observables, and the scheme dependence.
In Section~\ref{sec:NonFacPow} we briefly discuss our approach to non-factorisable power corrections, which differs from the popular procedure \cite{Egede:2008uy} of multiplying each amplitude with a complex factor. In Section~\ref{sec:results} we present our
final results for binned $B\to K^*\mu^+\mu^-$ observables.  We conclude in Section~\ref{sec:Conclu}.
Appendix~\ref{app1} addresses the issue of long-distance $c\bar c$ loops 
proposing different tests of the mechanism advocated in ref.~\cite{zwicky} to explain the $B\to K^*\mu^+\mu^-$ anomaly within the SM. A specific $B\to K^*\mu^+\mu^-$ observable is discussed where the advocated charm-loop contribution cannot mimic New Physics below the $J/\psi$ resonance, whereas two other tests are proposed to distinguish between SM long-distance effects and NP short-distance contributions. 
Appendix~\ref{app3} summarises the factorisable perturbative corrections used in the renormalisation schemes considered for our study and Appendix~\ref{app2} collects SM predictions for other $B\to K^*\mu^+\mu^-$ observables of interest.

\section{Soft form factors}
\label{sec:LO}

The evaluation of matrix elements for the decay $B\to K^*\mu^+\mu^-$ involves seven non-perturbative form factors
$V,A_{0,1,2},T_{1,2,3}$ (see ref.~\cite{Beneke:2000wa} for definitions). LCSR calculations of these form factors suffer
from large uncertainties originating from hadronic parameters, and moreover rely on certain assumptions (modelling the continuum contribution, fixing the Borel
parameter, etc.) introducing systematic uncertainties that are difficult to quantify. For a precise
analysis of the decay $B\to K^*\mu^+\mu^-$ it is thus desirable to reduce the sensitivity to the form factors as much as possible.
To this end one can make use of the fact that in the symmetry limit of large $K^*$ energies, i.e. for small invariant masses
$q^2$ of the lepton pair, the seven QCD form factors $V,A_{0,1,2},T_{1,2,3}$ reduce to two independent soft form factors
$\xi_{\perp,\parallel}$, up to corrections of order $\mathcal{O}(\alpha_s)$ and $\mathcal{O}(\Lambda/m_b)$. 
A completely general parametrisation for the QCD form factors $V,A_{1,2,0},T_{1,2,3}$ including all perturbative and
non-perturbative corrections is given by
\begin{eqnarray}
  V(q^2)&=&\frac{m_B+m_{K^*}}{m_B}\,\xi_{\perp}(q^2)\,+\,\Delta V^{\alpha_s}(q^2)\,+\,
           \Delta V^{\Lambda}(q^2)\,,\nonumber\\[2ex]
  A_1(q^2)&=&\frac{2E}{m_B+m_{K^*}}\,\xi_{\perp}(q^2)\,+\,\Delta A_1^{\alpha_s}(q^2)\,+\,
           \Delta A_1^{\Lambda}(q^2)\,,\nonumber\\[2ex]
  A_2(q^2)&=&\frac{m_B}{m_B-m_{K^*}}\,\left[\xi_{\perp}(q^2)-\xi_{\parallel}(q^2)\right]
            \,+\,\Delta A_2^{\alpha_s}(q^2)\,+
            \,\Delta A_2^{\Lambda}(q^2)\,,\nonumber\\[2ex]
  \label{eq:FF}
  A_0(q^2)&=&\frac{E}{m_{K^*}}\,\xi_{\parallel}(q^2)\,+\,\Delta A_0^{\alpha_s}(q^2)\,+\,
           \Delta A_0^{\Lambda}(q^2)\,,\\[2ex]
  T_1(q^2)&=&\xi_{\perp}(q^2)\,+\,\Delta T_1^{\alpha_s}(q^2)\,+\,
           \Delta T_1^{\Lambda}(q^2)\,,\nonumber\\[2ex]
  T_2(q^2)&=&\frac{2E}{m_B}\,\xi_{\perp}(q^2)\,+\,\Delta T_2^{\alpha_s}(q^2)\,+\,
           \Delta T_2^{\Lambda}(q^2)\,,\nonumber\\[2ex]
  T_3(q^2)&=&\left[\xi_{\perp}(q^2)-\xi_{\parallel}(q^2)\right]\,+\,\Delta T_3^{\alpha_s}(q^2)\,+\,
           \Delta T_3^{\Lambda}(q^2)\,,\nonumber
\end{eqnarray}
with $\Delta F^{\alpha_s}$ representing QCD corrections induced by hard gluons, and $\Delta F^{\Lambda}$ representing soft power
corrections of order $\mathcal{O}(\Lambda/m_b)$. Even though these corrections are expected to be small compared to the current
hadronic uncertainties of the QCD form factors, they play an important role in the study of optimised observables as they break the exact symmetry relations and therefore reintroduce a form factor dependence at order $\mathcal{O}(\alpha_s,\Lambda/m_b)$.
While QCD corrections $\Delta F^{\alpha_s}$ can be taken into account using results calculated within the framework of QCD
factorisation~\cite{Beneke:2000wa}, the inclusion of soft power corrections $\Delta F^{\Lambda}$ is not straightforward,
since no first-principle calculation of these quantities exists.

On the other hand, LCSR determinations of the QCD form factors $V,A_{1,2,0},T_{1,2,3}$ include all factorisable power
corrections. Therefore as long as one is not interested in an explicit decomposition of the form factors into a soft
contribution and power corrections, one can directly use the LCSR results as input for the form factors appearing in the naively
factorised expressions for the amplitudes. In order to obtain precise predictions for observables involving QCD form factors, it is 
essential to assess properly all correlations among the errors of the different form factors within the LCSR calculation.
The decomposition (\ref{eq:FF}), on the other hand, if supplemented by a realistic estimate regarding the size of the
$\mathcal{O}(\Lambda/m_b)$ corrections $\Delta F^{\Lambda}$, takes into account the major part of correlations among the form
factors by  representing them in terms of the two soft form factors $\xi_{\perp,\parallel}$. Therefore as long as the correlations
among the LCSR form factors are not accessible \emph{or} are not known to the same degree as they can be inferred from
eq.~(\ref{eq:FF}), making use of the soft form factor decomposition is very convenient in order to obtain precise results
for angular observables in the decay $B\to K^*\mu^+\mu^-$.

The separation of the form factors $V,A_{1,2,0},T_{1,2,3}$ into soft form factors $\xi_{\perp,\parallel}$ and
perturbative/power
corrections $\Delta F^{\alpha_s,\Lambda}$ in eq.~(\ref{eq:FF}) is not unique as one can always redefine
$\xi_{\perp,\parallel}$ in such a way that these corrections are partly absorbed. In order to unambiguously define the soft
form factors $\xi_{\perp,\parallel}$ (and thus the terms $\Delta F^{\alpha_s,\Lambda}$ ), one first has to
fix a renormalisation scheme, i.e. define the $\xi_{\perp,\parallel}$ in terms of the physical form factors
$V,A_{1,2,0},T_{1,2,3}$. 

A popular definition for $\xi_{\perp}$, used for example in refs.~\cite{Beneke:2000wa,Altmannshofer:2008dz,Descotes-Genon:2013vna},
is
\begin{equation}
  \xi^{(1)}_{\perp}(q^2)\,\equiv\,\frac{m_B}{m_B+m_{K^*}}V(q^2).
  \label{eq:xiV}
\end{equation}
where the superscript refers to the scheme thus defined.
This definition eliminates all corrections to the form factor $V$ leading to $\Delta V^{\alpha_s}(q^2)=\Delta V^{\Lambda}(q^2)=0$.
Alternatively one can define a second scheme for $\xi_{\perp}$, in terms of $T_1$,
\begin{equation}
  \xi^{(2)}_{\perp}(q^2)\,\equiv\,T_1(q^2),
  \label{eq:xiT1}
\end{equation}
eliminating in this way $\Delta T_1^{\alpha_s}(q^2),\Delta T_1^{\Lambda}(q^2)$. This choice of scheme has been applied in
refs.~\cite{Beneke:2004dp,Jager:2012uw}, being quite convenient when extracting $T_1(0)$ from experimental
data on $B\to K^*\gamma$. Note, however, that extracting $T_1(0)$ from $B\to K^*\gamma$ relies on the assumption that there is no
new physics in the Wilson coefficients $C_7$ and $C_7^{\prime}$. Furthermore, the $T_1^{\textrm{exp}}(0)$ determined in this way can be
identified with the form factor $T_1(0)$ only up to corrections of order $\mathcal{O}(\Lambda/m_b)$, 
stemming from four-quark operators (e.g., $c\bar c$ loops). These non-factorisable power corrections can neither be
computed nor extracted from the QCD factorisation prediction for $B\to K^*\gamma$. Therefore, identifying $T_1^\text{exp}(0)$ with
$T_1(0)$ amounts to including unknown non-factorisable power corrections into $T_1$.
Hence it cannot be used consistently as input in our approach to determine the
factorisable $\mathcal{O}(\Lambda/m_b)$-corrections, and we will instead infer $T_1$ from LCSR calculations.

The soft form factor $\xi_{\parallel}$ can be defined as
\begin{equation}
  \xi^{(1)}_{\parallel}(q^2)\,\equiv\,\frac{m_B+m_{K^*}}{2E}A_1(q^2)\,-\,\frac{m_B-m_{K^*}}{m_B}A_2(q^2),
  \label{eq:xiA12}
\end{equation}
as done for example in Refs.~\cite{Beneke:2004dp,Altmannshofer:2008dz,Descotes-Genon:2013vna}. This definition
minimises power corrections in the form factors $A_{1,2}$ by correlating $\Delta A_1^{\alpha_s}(q^2),\Delta A_1^{\Lambda}(q^2)$
with $\Delta A_2^{\alpha_s}(q^2)$ and $\Delta A_2^{\Lambda}(q^2)$.  An alternative scheme applied in
ref.~\cite{Beneke:2000wa} is given by
\begin{equation}
  \xi^{(2)}_{\parallel}(q^2)\equiv \frac{m_{K^*}}{E}A_0(q^2).
  \label{eq:xiA0}
\end{equation}

The choice of scheme determines which part of the $\mathcal{O}(\alpha_s,\Lambda/m_b)$ corrections will be absorbed into
$\xi_{\perp,\parallel}$ and which part will remain in the functions $\Delta F^{\alpha_s,\Lambda}$. 
The perturbative corrections $\Delta F^{\alpha_s}$ can be computed explicitly in each scheme, as illustrated in App.~\ref{app3}.
If one had full control on the power corrections $\Delta F^{\Lambda}$
(including correlations among their errors), 
physical quantities would not depend on the choice of scheme for the soft form
factors at $\mathcal{O}(\alpha_s)$.\footnote{
There is still a small residual scheme dependence at $\mathcal{O}(\Lambda/m_b)$
introduced by non-factorisable power corrections.
} On the other hand, as long as information on the $\Delta F^{\Lambda}$ is not available or only available in part (for example
because correlations cannot be assessed), predictions for observables will exhibit a scheme dependence at
$\mathcal{O}(\Lambda/m_b)$. In this situation a proper choice of scheme can increase the precision of the theoretical prediction. 
Assume for example that a certain observable is dominated by the form factor $V$. Obviously a prediction employing scheme 1
for $\xi_{\perp}$ where $V$ is directly taken as input will be more accurate in this case than a prediction relying on scheme 2
where $V$ is obtained as a sum of $T_1$ and an unknown (or only partially known) power correction $\Delta V^{\Lambda}$.
This also depends on the relative size of the LCSR uncertainties in $V$ and $T_1$. If $T_1$ is known much more precisely, and
the total uncertainty in $V$ is larger than expected power corrections, scheme 2 might be preferred in this case.
The general statement is the following: Different schemes lead to different uncertainties, and for each observable there is a
preferred scheme where uncertainties are minimised.\footnote{Of course, for some observables different schemes might
lead to very similar uncertainties; in these cases the choice of scheme has no impact.}

Different choices of the renormalisation scheme correspond to a reshuffling between soft form factors and power corrections. 
This choice affects the pattern of cancellation of power corrections when one considers clean observables. Indeed, since
the soft form factors $\xi_{\perp,\parallel}$ cancel at leading order in clean observables, any power correction absorbed into the soft form factors according to the chosen renormalisation scheme will undergo a similar cancellation, so that it can contribute only at order $\mathcal{O}(\alpha_s,\Lambda/m_b)\times\mathcal{O}(\Lambda/m_b)$ (the second factor coming from the power correction itself). On the other hand,
the power corrections that are kept explicitly in $\Delta F^{\Lambda}$ contribute at $\mathcal{O}(\Lambda/m_b)$ and their size must be assessed. Therefore the choice of the renormalisation scheme is crucial when one wants to determine how power corrections will affect
clean observables.

\begin{figure}
\includegraphics[width=7.5cm]{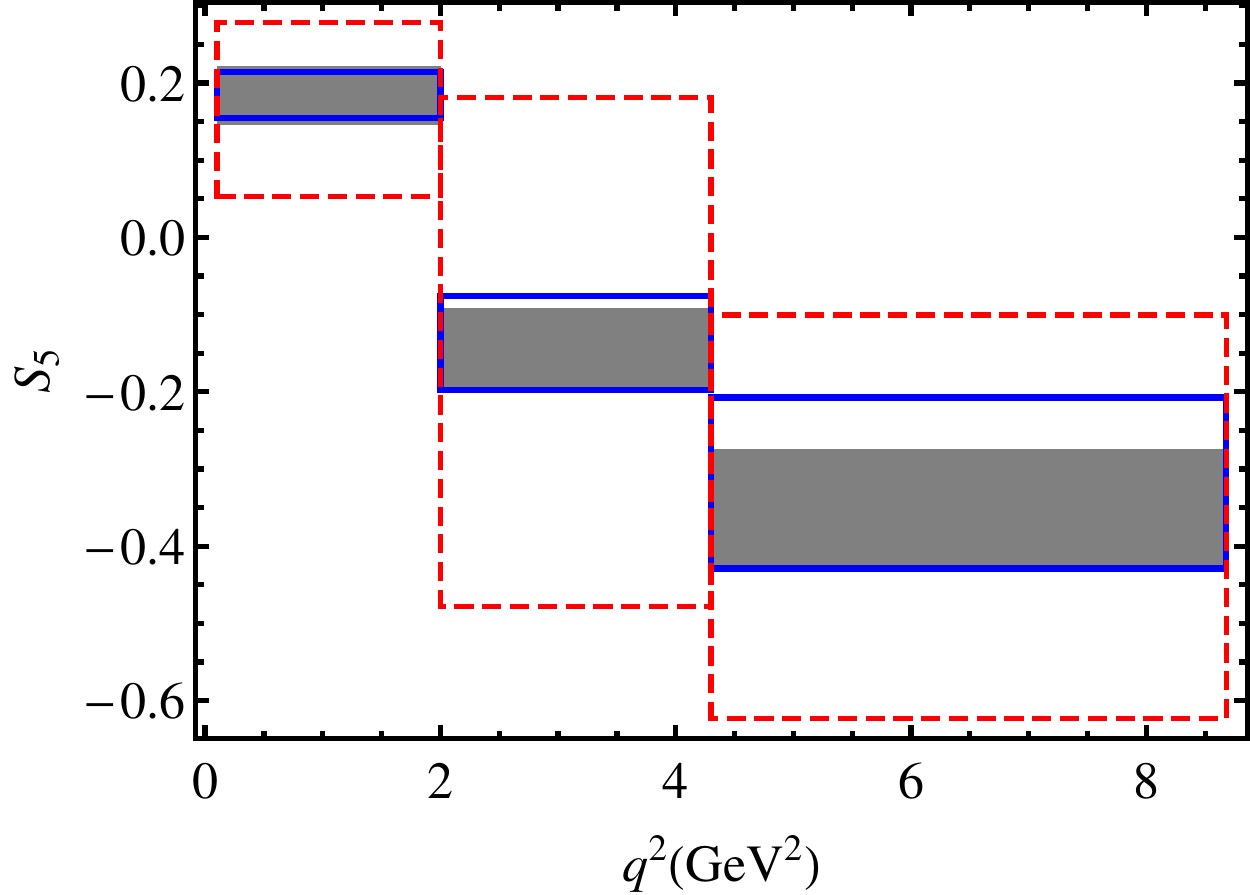}\hspace{7mm}
\includegraphics[width=7.5cm]{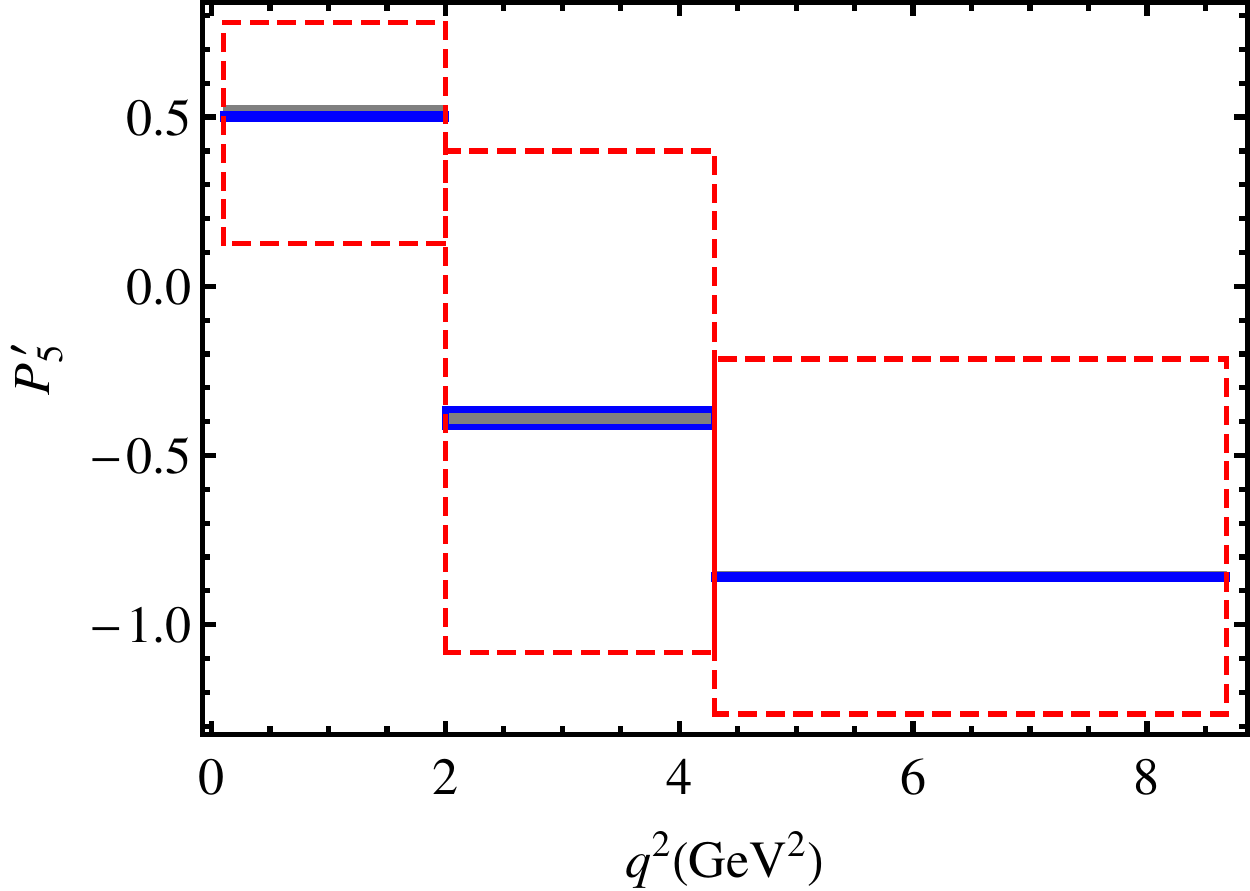}
\caption{Scheme dependence in the prediction of the observables $S_5$ and $P_5'$. Power corrections are set to zero and
uncertainties are solely due to form factors. Gray bands correspond to scheme 1 ($V,A_1,A_2$), blue (solid) boxes to
scheme 2 ($T_1,A_0$), and red (dashed) boxes to the full-form-factor approach \emph{with no correlations}. Form factor input
is taken from ref.~\cite{Khodjamirian:2010vf} in all cases.}
\label{fig:LO}
\end{figure}

In fig~\ref{fig:LO} we show the leading-power predictions for the observable $S_5$~\cite{Altmannshofer:2008dz},
and the optimised observable $P_5^{\prime}$~\cite{DescotesGenon:2012zf} in two different schemes:
$(\xi_{\perp}^{(1)},\xi_{\parallel}^{(1)})$
defined from $(V,A_1,A_2)$, and $(\xi_{\perp}^{(2)},\xi_{\parallel}^{(2)})$ defined from $(T_1,A_0)$. As input
we have used the LCSR form factors from ref.~\cite{Khodjamirian:2010vf}. The observable $S_5$ indeed exhibits the expected scheme
dependence which can be considered as a measure of $\mathcal{O}(\Lambda/m_b)$ power corrections. The observable $P_5^{\prime}$,
on the other hand, shows only a very mild scheme dependence because the soft form factors cancel at leading order pushing the
scheme dependence to $\mathcal{O}(\alpha_s)$. In addition we show the prediction which one would obtain using uncorrelated
QCD form factors without resorting to the soft form factor decomposition.

\section{Factorisable power corrections}
\label{sec:FacPow}

Even though no direct calculation of the factorisable power corrections $\Delta F^{\Lambda}$ exists, the fact that they are
included in LCSR determinations of the QCD form factors allows for their estimation. One studies to which
extend the full LCSR form factors violate the ($\Delta F^{\alpha_s}$-corrected) symmetry relations~(\ref{eq:FF}) and attributes
these deviations to the $\Delta F^{\Lambda}$, which then can be determined from a fit. This basic strategy has been proposed and applied for the first time in ref.~\cite{Jager:2012uw}. In our analysis we modify the approach of ref.~\cite{Jager:2012uw}
and go beyond it in several aspects. In the following we will specify our method in detail pointing out 
the differences with respect to ref.~\cite{Jager:2012uw}.

\subsection{General approach}
\label{sec:genap}

Following ref.~\cite{Jager:2012uw} we parametrise the unknown soft power corrections $\Delta F^{\Lambda}$ as a polynomial
in $q^2/m_B^2$,
\begin{equation}\label{eq:PCparaJager}
   \Delta F^{\Lambda}(q^2) = a_F\,+\,b_F\,\frac{q^2}{m_B^2}\,+\,c_F\,\frac{q^4}{m_B^4}\,+\,\ldots,
\end{equation}
and perform a fit of the resulting form factor representation (\ref{eq:FF}) to the QCD form factors from LCSR, using central
values for the latter. In ref.~\cite{Jager:2012uw} this fit was performed to first order in $q^2/m_B^2$ and the result
$\hat{a}_F,\hat{b}_F$ was interpreted as an order-of-magnitude estimate for power corrections. Consequently the error
associated to factorisable power corrections was estimated by varying independently
$-|\hat{a}_F|\leq a_F\leq +|\hat{a}_F|$, $-|\hat{b}_F|\leq b_F\leq |\hat{b}_F|$ assuming the central values of $\Delta F^{\Lambda}$ to be zero.
In our analysis we perform a fit to second order in $q^2/m_B^2$ and keep the correlated results
$\hat{a}_F,\hat{b}_F,\hat{c}_F$ as (non-zero) central values for $\Delta F^{\Lambda}$. With this procedure the central values of
our predictions of observables will agree exactly with the ones which one would obtain in a calculation based on full LCSR form
factors. In particular, they will not exhibit any dependence on the renormalisation
scheme chosen for the soft form factors $\xi_{\perp,\parallel}$, apart from the one induced by non-factorisable power
corrections.

\begin{table}
\footnotesize
\centering
\begin{tabular}{@{}l|rrr|ccc@{}}
\hline\hline 
& $\hat{a}_F^{(1)}\qquad$ & $\hat{b}_F^{(1)}\qquad$ & $\hat{c}_F^{(1)}\qquad$ &$r(0\,\rm{GeV}^2)$ & $r(4\,\rm{GeV}^2)$ & $r(8\,\rm{GeV}^2)$ \\ 
\hline 
$A_0$(KMPW)& $0.002\pm0.000$ & $0.590\pm 0.125$ & $1.473\pm 0.251$ & $0.007$ & $0.220$ & $0.333$ \\ 
$A_0$(BZ)& $0.000\pm 0.000$ & $0.003\pm 0.052$ & $0.219\pm 0.121$ & $0.002$ & $0.012$ & $0.032$ \\ 
\hline 
$A_1$(KMPW)& $-0.013\pm 0.025$ & $-0.056\pm 0.018$ & $0.158\pm 0.021$ & $0.052$ & $0.063$ & $0.049$ \\ 
$A_1$(BZ)& $-0.009\pm 0.027$ & $0.042\pm 0.018$ & $0.078\pm 0.017$ & $0.032$ & $0.003$ & $0.029$ \\ 
\hline 
$A_2$(KMPW)& $-0.018\pm 0.023$ & $-0.105\pm 0.022$ & $0.192\pm 0.028$ & $0.078$ & $0.108$ & $0.101$ \\ 
$A_2$(BZ)& $-0.012\pm 0.024$ & $0.037\pm 0.029$ & $0.239\pm 0.034$ & $0.050$ & $0.006$ & $0.053$ \\ 
\hline 
$T_1$(KMPW)& $-0.006\pm 0.031$ & $-0.012\pm 0.054$ & $-0.034\pm 0.095$ & $0.016$ & $0.018$ & $0.020$ \\ 
$T_1$(BZ)& $-0.024\pm 0.032$ & $-0.019\pm 0.045$ & $-0.014\pm 0.092$ & $0.075$ & $0.066$ & $0.057$ \\ 
\hline 
$T_2$(KMPW)& $-0.005\pm 0.031$ & $0.153\pm 0.043$ & $0.544\pm 0.061$ & $0.014$ & $0.075$ & $0.174$ \\ 
$T_2$(BZ)& $-0.024\pm 0.031$ & $0.040\pm 0.021$ & $0.072\pm 0.019$ & $0.074$ & $0.046$ & $0.015$ \\ 
\hline 
$T_3$(KMPW)& $-0.002\pm 0.022$ & $0.308\pm 0.059$ & $0.786\pm 0.093$ & $0.007$ & $0.181$ & $0.322$ \\ 
$T_3$(BZ)& $-0.035\pm 0.019$ & $-0.021\pm 0.021$ & $0.097\pm 0.025$ & $0.178$ & $0.154$ & $0.116$ \\ 
\hline 
\hline 
\end{tabular}
\caption{Fit results for the power-correction parameters in the case of scheme 1 --with $(\xi_{\perp}^{(1)},\xi_{\parallel}^{(1)})$
defined from $(V,A_1,A_2)$. The relative size $r(q^2)$ is also shown for $q^2=0\,\rm{GeV}^2,4\,\rm{GeV}^2,8\,\rm{GeV}^2$. The label
KMPW refers to LCSR input from ref.~\cite{Khodjamirian:2010vf}, and BZ to ref.~\cite{Ball:2004ye}.
In this scheme, $V$ receives no power corrections and therefore the corresponding parameters vanish.}
\label{tab:fit1}
\end{table}

For the error estimate we vary $a_F,b_F,c_F$ symmetrically around their respective central values:
\eqa{
   \hat{a}_F-\Delta \hat{a}_F\leq &a_F&\leq \hat{a}_F+\Delta\hat{a}_F\ ,\nn\\
   \hat{b}_F-\Delta \hat{b}_F\leq &b_F&\leq \hat{b}_F+\Delta\hat{b}_F,\ \\
   \hat{c}_F-\Delta \hat{c}_F\leq &c_F&\leq \hat{c}_F+\Delta\hat{c}_F\ .\nn
}
In principle the errors $\Delta\hat{a}_F,\Delta\hat{b}_F,\Delta\hat{c}_F$ are related to the errors of the QCD form
factors and could be determined from a fit if the correlations among the form factors were known precisely. In the absence
of such knowledge one is forced to rely on dimensional arguments, 
exploiting the $\Lambda/m_b$ suppression of
the $\Delta F^{\Lambda}$. To this end we consider an expanded approximation $F(q^2)=A_F+B_Fq^2/m_B^2+C_F q^4/m_B^4$ of the full LCSR form factors and attribute a 10\% error to the power corrections setting
$\Delta\hat{a}_F=0.1A_F,\Delta\hat{b}_F=0.1B_F,\Delta\hat{c}_F=0.1C_F$~\footnote{The expanded approximation is only
used to obtain a normalisation for the errors $\Delta\hat{a}_F,\Delta\hat{b}_F,\Delta\hat{c}_F$, while everywhere else in our analysis the full
$q^2$-dependence of the form factors is used.}. Given the fact that
$\Delta F^{\Lambda}\sim F\times \mathcal{O}(\Lambda/m_b)\sim 0.1F$ this amounts to assigning an error of $\sim 100\%$ to the 
result $\Delta\hat{F}$ from the fit. 

Note that with our approach any future improvement on the precision of form factor
calculations can be accounted for by reducing the size of the free parameters $\Delta\hat{a}_F,\Delta\hat{b}_F,\Delta\hat{c}_F$
accordingly. On the contrary, in the method of ref.~\cite{Jager:2012uw} the errors are frozen due to their determination from central values, and they do not approach zero in the hypothetical limit of exact knowledge of the form factors, if (as expected) they
do not fulfill exactly the leading power symmetry relations.

The soft form factor decomposition~(\ref{eq:FF}) is not unique and depends on the renormalisation scheme
for the soft form factors $\xi_{\perp},\xi_{\parallel}$. In the following section \ref{sec:schemedep} we will discuss 
how the choice of scheme affects the errors induced by power corrections for $B\to K^*\mu^+\mu^-$ angular observables.

\begin{table}
\footnotesize
\centering
\begin{tabular}{@{}l|rrr|ccc@{}}
\hline\hline 
& $\hat{a}_F^{(2)}\qquad$ & $\hat{b}_F^{(2)}\qquad$ & $\hat{c}_F^{(2)}\qquad$ &
$r(0\,\rm{GeV}^2)$ & $r(4\,\rm{GeV}^2)$ & $r(8\,\rm{GeV}^2)$ \\ 
\hline 
$V$(KMPW)& $0.005\pm 0.036$ & $0.013\pm 0.063$ & $0.039\pm 0.113$ & $0.016$ & $0.018$ & $0.020$ \\ 
$V$(BZ)& $0.027\pm 0.039$ & $0.021\pm 0.053$ & $0.014\pm 0.107$ & $0.072$ & $0.064$ & $0.056$ \\ 
\hline 
$A_1$(KMPW)& $-0.009\pm 0.025$ & $-0.049\pm 0.018$ & $0.166\pm 0.021$ & $0.035$ & $0.043$ & $0.027$ \\ 
$A_1$(BZ)& $0.011\pm 0.027$ & $0.038\pm 0.018$ & $0.069\pm 0.017$ & $0.043$ & $0.061$ & $0.083$ \\ 
\hline 
$A_2$(KMPW)& $-0.010\pm 0.023$ & $0.099\pm 0.022$ & $1.496\pm 0.028$ & $0.040$ & $0.135$ & $0.451$ \\ 
$A_2$(BZ)& $0.017\pm 0.024$ & $0.055\pm 0.029$ & $0.400\pm 0.034$ & $0.071$ & $0.115$ & $0.187$ \\ 
\hline 
$T_2$(KMPW)& $0.000\pm 0.000$ & $0.161\pm 0.043$ & $0.553\pm 0.061$ & $0.002$ & $0.092$ & $0.191$ \\ 
$T_2$(BZ)& $0.000\pm 0.000$ & $0.035\pm 0.021$ & $0.062\pm 0.019$ & $0.000$ & $0.019$ & $0.040$ \\ 
\hline 
$T_3$(KMPW)& $0.005\pm 0.022$ & $0.486\pm 0.059$ & $1.895\pm 0.093$ & $0.026$ & $0.352$ & $0.639$ \\ 
$T_3$(BZ)& $-0.011\pm 0.019$ & $-0.006\pm 0.021$ & $0.235\pm 0.025$ & $0.054$ & $0.028$ & $0.027$ \\ 
\hline 
\hline 
\end{tabular}
\caption{Fit results for the power-correction parameters in the case of scheme 2 --with $(\xi_{\perp}^{(2)},\xi_{\parallel}^{(2)})$
defined from $(T_1,A_0)$. The relative size $r(q^2)$ is also shown for $q^2=0\,\rm{GeV}^2,4\,\rm{GeV}^2,8\,\rm{GeV}^2$.
The label KMPW refers to LCSR input from ref.~\cite{Khodjamirian:2010vf}, and BZ to ref.~\cite{Ball:2004ye}.
In this scheme, $A_0$ and $T_1$ receive no power corrections and therefore the corresponding parameters vanish.}
\label{tab:fit2}
\end{table}

In tables~\ref{tab:fit1} and \ref{tab:fit2} we show respectively our fit results in the two different schemes, with
$(\xi_{\perp}^{(1)},\xi_{\parallel}^{(1)})$ defined from $(V,A_1,A_2)$ and with
$(\xi_{\perp}^{(2)},\xi_{\parallel}^{(2)})$ defined from $(T_1,A_0)$, and for two different sets of LCSR form factors \cite{Khodjamirian:2010vf,Ball:2004ye}. Apart from the actual values of
the coefficients $\hat{a}_F$, $\hat{b}_F$, $\hat{c}_F$ and the estimated errors, we also display the relative size
\begin{equation}
   r(q^2)\,=\,\left|\frac{\hat{a}_F+\hat{b}_F\frac{q^2}{m_B^2}+\hat{c}_F\frac{q^4}{m_B^4}}{F(q^2)}\right|
\end{equation}
for different invariant masses $q^2=0\,\rm{GeV}^2,4\,\rm{GeV}^2,8\,\rm{GeV}^2$ of the lepton pair.
The results confirm that power corrections are typically $\lesssim 10\%$ for $q^2\leq 4$ GeV$^2$ as expected from dimensional arguments.
In the case of LCSR input from ref.~\cite{Khodjamirian:2010vf} (KMPW) slightly larger power corrections are found
for larger values of $q^2$ for the form factor $T_3$, as well as for $A_0$ in scheme 1 ($A_2$ in scheme 2).
However, this is not problematic in the case of the $B\to K^*\mu^+\mu^-$ transversity amplitudes, given that $A_0$ is suppressed by powers of the lepton mass and $T_3$ is relatively subdominant as compared to other tensor contributions due to their relative kinematic prefactors at large recoil~\cite{Descotes-Genon:2013vna}.

\subsection{Correlations of power corrections}
\label{sec:schemedep}

The quantities $a_F,b_F,c_F$ parametrising the factorisable power corrections are subject to several constraints, resulting from
(a)  kinematic correlations among QCD form factors at maximum recoil, and (b) the definition of the soft form factors
$\xi_{\perp}$ and $\xi_{\parallel}$. Taking into account these correlations reduces the number of parameters to be varied in the error analysis, reducing correspondingly the overall uncertainties in the observables. Not taking into account such correlations would lead to an over-estimation of the effect of factorisable power corrections.

At $q^2=0$ the QCD form factors obey the exact equations\footnote{The relation between $A_0,A_1$ and $A_2$ is only approximately fulfilled for the input from LCSR determinations. In practice we enforce it to hold exactly by a 
rescaling of $A_0$.}
\begin{eqnarray}
   A_0(0)&=&\frac{m_B+m_{K^*}}{2m_{K^*}}\,A_1(0)\,-\,\frac{m_B-m_{K^*}}{2m_{K^*}}\,A_2(0)\,, \nonumber\\[1ex]
   T_1(0)&=&T_2(0)\,.\label{eq:q20rel}
\end{eqnarray}
These equations imply that the soft power corrections fulfil
\begin{eqnarray}
  a_{A_0}&=&\frac{m_B+m_{K^*}}{2m_{K^*}}\,a_{A_1}\,-\,\frac{m_B-m_{K^*}}{2m_{K^*}}\,a_{A_2}\,\,,\nonumber\\[1ex] a_{T_1}&=&
  a_{T_2}\,\,.\label{eq:consistency2}
\end{eqnarray}

While the correlations of eq.~(\ref{eq:consistency2}) always apply, additional constraints depend on the renormalisation
scheme chosen for $\xi_{\perp}$ and $\xi_{\parallel}$.
Defining $\xi_{\perp}$ in terms of $V$ according to eq.~(\ref{eq:xiV}) results in
\begin{equation}\label{eq:Vconstraint}
   a^{(1)}_V=0,\hspace{2cm}b^{(1)}_V=0,\hspace{2cm}c^{(1)}_V=0,
\end{equation}
while a definition from $T_1$ following eq.~(\ref{eq:xiT1}) gives
\begin{equation}
   a^{(2)}_{T_1}=0,\hspace{2cm}b^{(2)}_{T_1}=0,\hspace{2cm}c^{(2)}_{T_1}=0.
\end{equation}
If the soft form factor $\xi_{\parallel}$ is defined from $A_{1,2}$ in eq.~(\ref{eq:xiA12}), one finds the correlations
\begin{eqnarray}
   a^{(1)}_{A_2}&=&\frac{m_B+m_{K^*}}{m_B-m_{K^*}}a^{(1)}_{A_1}\,,\nonumber\\
   b^{(1)}_{A_2}&=&\frac{m_B+m_{K^*}}{m_B-m_{K^*}}\left[a^{(1)}_{A_1}+b^{(1)}_{A_1}\right]\nonumber\\
   c^{(1)}_{A_2}&=&\frac{m_B+m_{K^*}}{m_B-m_{K^*}}\left[a^{(1)}_{A_1}+b^{(1)}_{A_1}
   +c^{(1)}_{A_1}\right]\   \label{eq:A12constraint}
\end{eqnarray}
for the corresponding power corrections. The definition (\ref{eq:xiA0}) in terms of $A_0$, on the other hand, translates into
\begin{equation}
   a^{(2)}_{A_0}=0,\hspace{2cm}b^{(2)}_{A_0}=0,\hspace{2cm}c^{(2)}_{A_0}=0.
   \label{eq:A0constraint}
\end{equation}

Note that unlike the authors of ref.~\cite{Jager:2012uw}, we do not enforce any of the constraints (either the general constraints eqs.~(\ref{eq:q20rel})-(\ref{eq:consistency2}) or the renormalisation-scheme dependent ones eqs.~(\ref{eq:Vconstraint})-(\ref{eq:A12constraint})) in the fit for the central values $\hat{a}_F,\hat{b}_F,\hat{c}_F$. Our results from the fit given in tables~\ref{tab:fit1} and \ref{tab:fit2} respect the constraints within the overall accuracy of the fit, limited by the parametrisation of the power correction functions as second order polynomials. The precision to which the correlations are fulfilled can be improved by adding higher-order coefficients $d_F,e_F,...$ in the fit~\footnote{Imposing the correlations in the fit by hand would not improve the overall accuracy of the fit result. As constraints are mostly related to the endpoint $q^2=0$, it would imply that form factor values $F(0)$ at $q^2=0$ have a larger weight in the fit than $F(q^2)$ at larger $q^2$. The resulting functions for the sum of soft form factors and power corrections would describe then the full form factors better at $q^2\approx 0$, but worse in the physically more interesting region $q^2>1\,\rm{GeV}^2$.}.

\begin{figure}
\centering
\includegraphics[width=7cm,height=4.6cm]{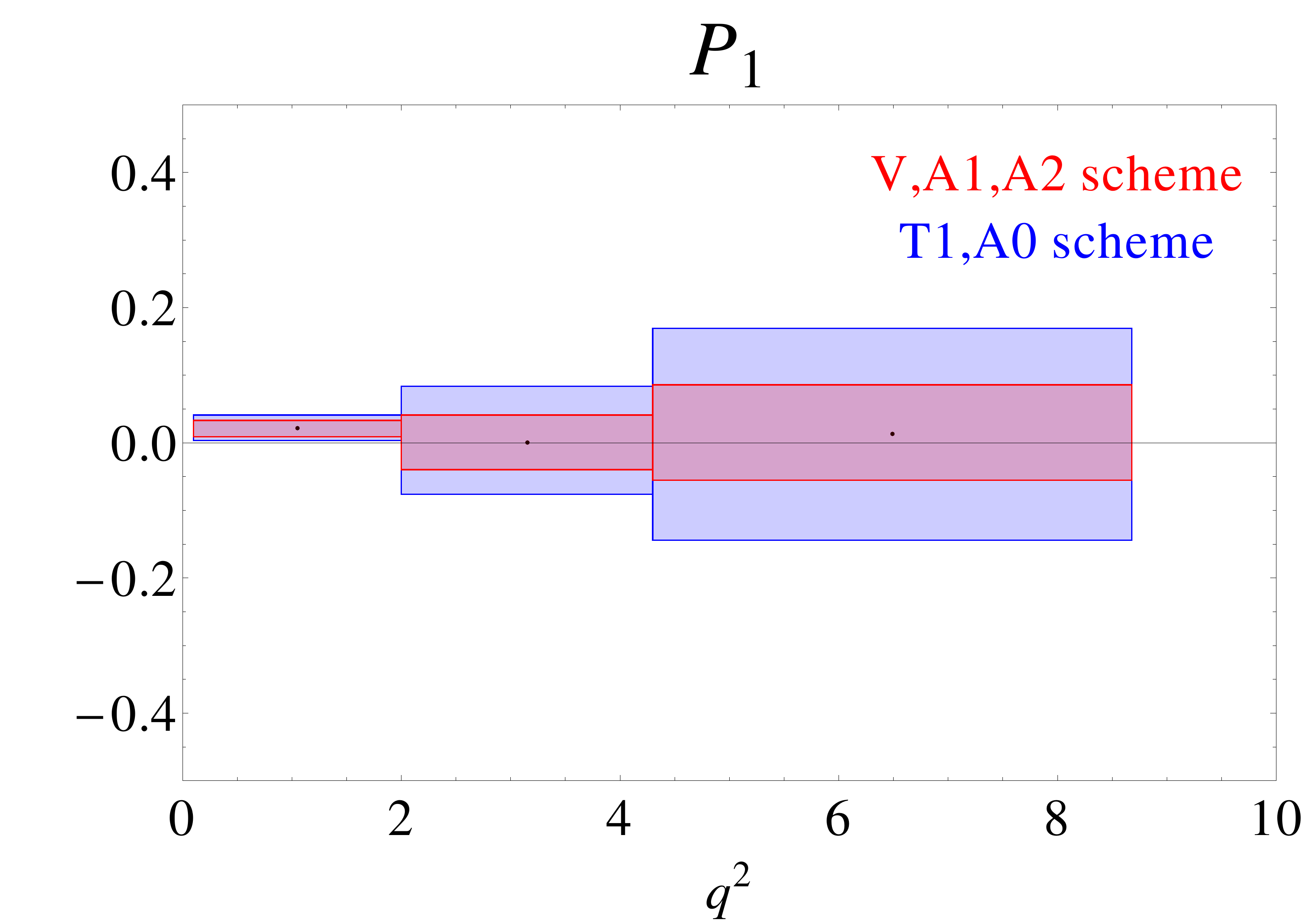}\hspace{1cm}
\includegraphics[width=7cm,height=4.6cm]{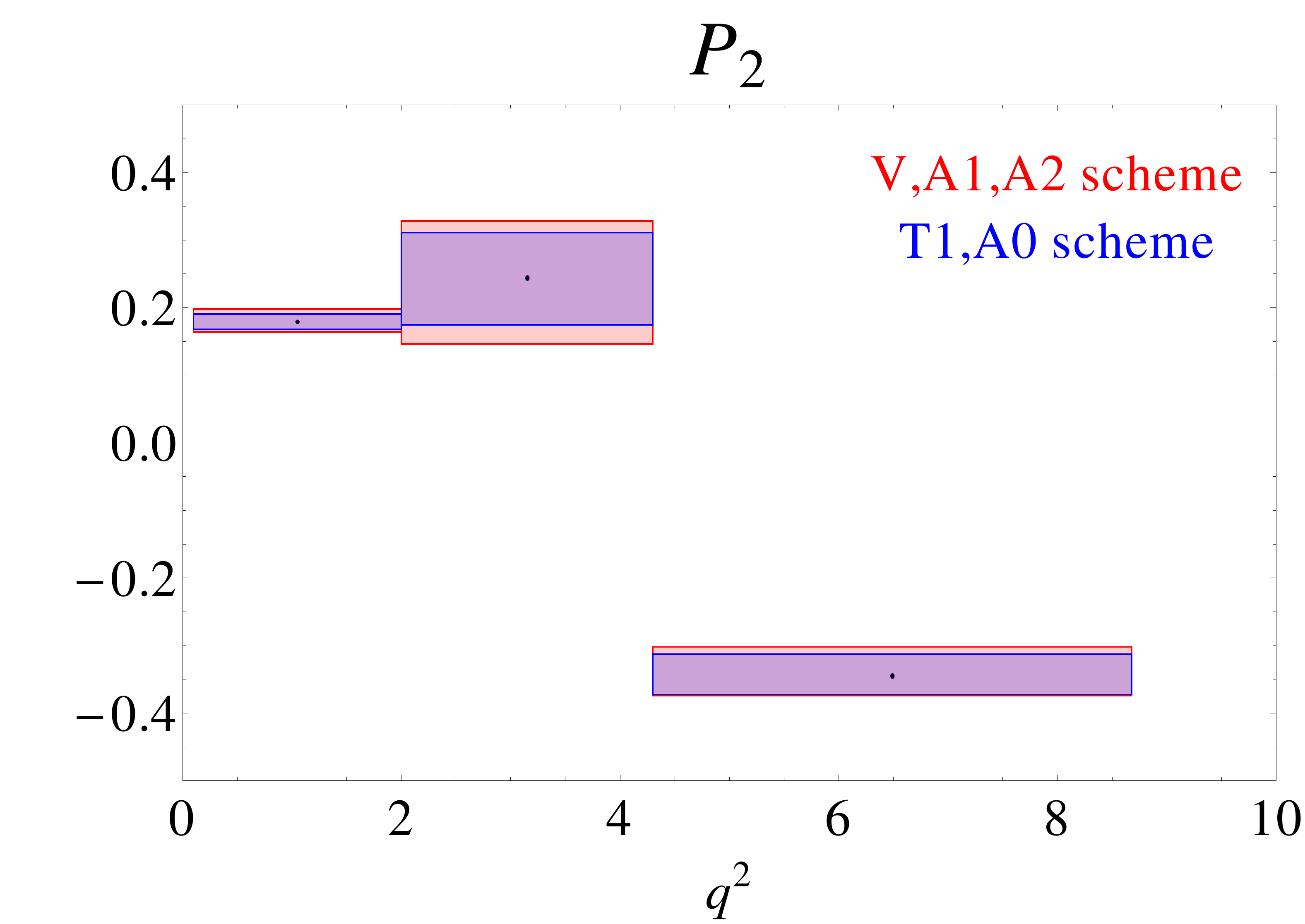}\\[2mm]
\includegraphics[width=7cm,height=4.6cm]{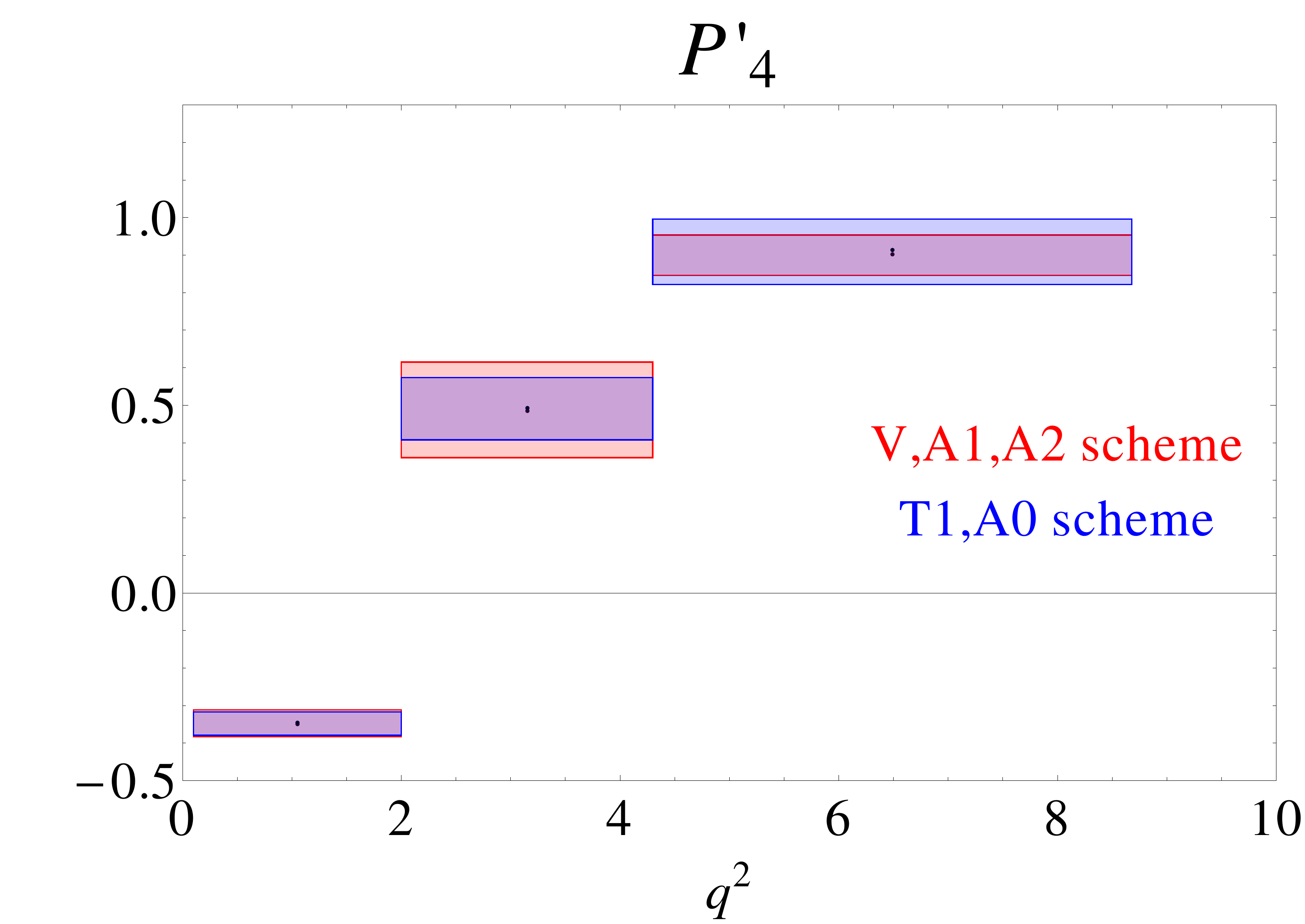}\hspace{1cm}
\includegraphics[width=7cm,height=4.6cm]{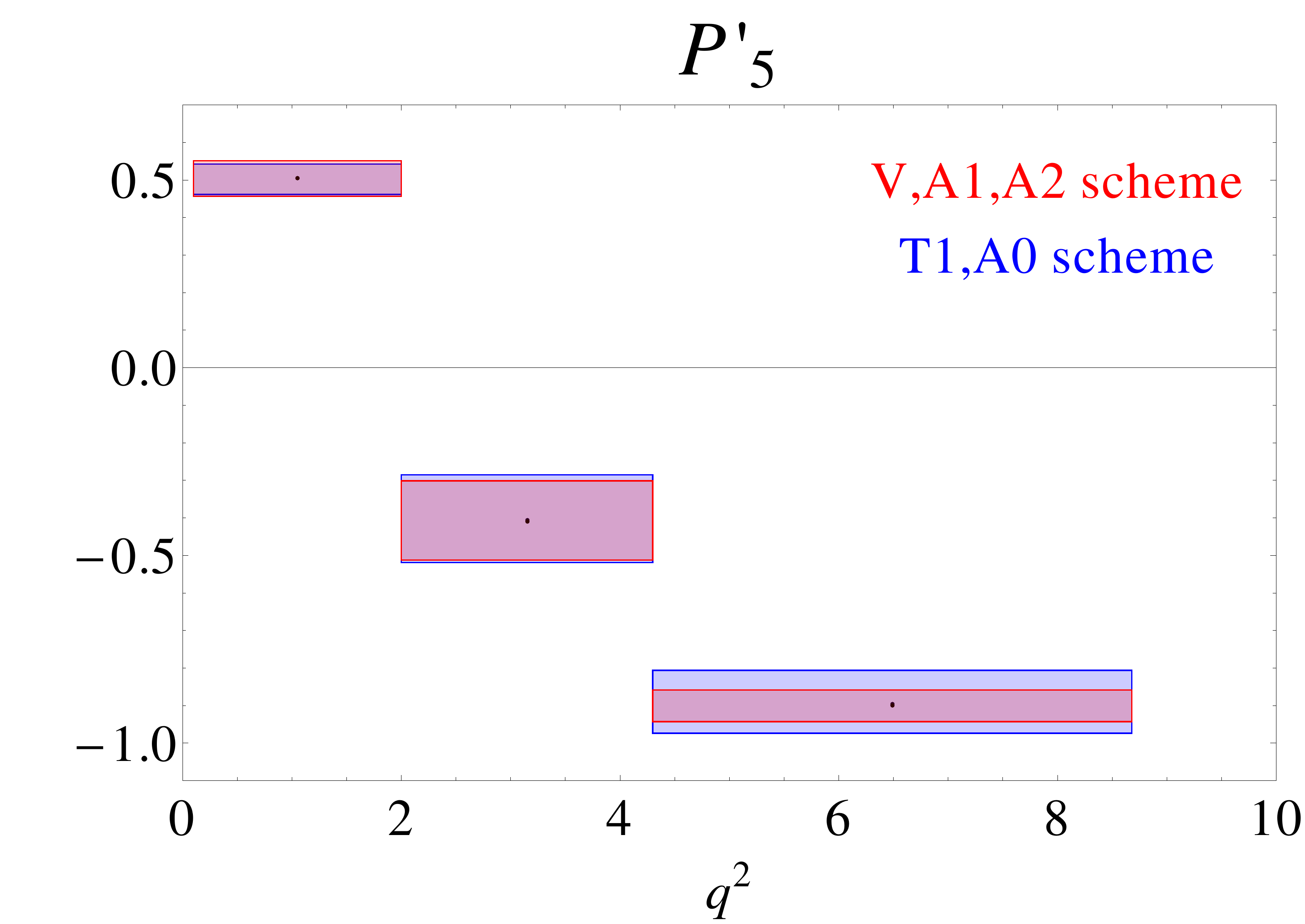}
\caption{Scheme dependence on the prediction of the observables $P_1$, $P_2$, $P_4'$, $P_5'$ in QCD factorisation. These results include factorisable
power corrections as described in the text.}
\label{fig_SD}
\end{figure}

For the estimation of errors associated to power corrections, we vary the parameters $a_F,b_F,c_F$ within the ranges specified in tables~\ref{tab:fit1} and \ref{tab:fit2}, imposing in addition the constraints (\ref{eq:consistency2})--(\ref{eq:A0constraint}) according to the respective scheme. As the correlations depend on the definition chosen for the soft form factors $\xi_{\perp,\parallel}$, the errors originating from factorisable power corrections are scheme dependent.
In Figure~\ref{fig_SD} we show the corresponding errors for the observables $P_1$, $P_2$, $P_4^{\prime}$ and $P_5^{\prime}$ in the two schemes, with $(\xi_{\perp}^{(1)},\xi_{\parallel}^{(1)})$ defined from $(V,A_1,A_2)$ and with $(\xi_{\perp}^{(2)},\xi_{\parallel}^{(2)})$ defined from $(T_1,A_0)$~\footnote{Obviously, these two examples are not limitative: other pairings of normalisation schemes could be considered, and additional schemes could be devised.}. As input we have used the LCSR form factors from ref.~\cite{Khodjamirian:2010vf}. For 
$q^2> 4\,\rm{GeV}^2$, the observables $P_1$ and $P_5^{\prime}$ exhibit significantly smaller errors in the first scheme, while the observables $P_2$ and $P_4^{\prime}$ have slightly smaller uncertainties in the second scheme. 

The scheme dependence of the observables is dominated by the definition of $\xi_{\perp}$. The fact that the Wilson
coefficient $C_9$ always enters in combination with a vector form factor $V,A_1,A_2$, while $C_7$ enters in combination
with a tensor form factor $T_{1,2,3}$, thus explains that observables with a high sensitivity to $C_9$ like the third bin
of $P_5^{\prime}$ can be predicted more precisely in the first scheme, while observables with a higher sensitivity to $C_7$
like $P_2$ are better described in the second. Concerning $\xi_{\parallel}$, the situation is unambiguous: since any contribution
of the form factor $A_0$ to physical observables is always suppressed by small lepton masses, the constraint
(\ref{eq:A0constraint}) does not improve the precision of the theory predictions in contrast to the correlation
(\ref{eq:A12constraint}) obtained in the first scheme. On the other hand, one may think that the first scheme has the
disadvantage of $\xi_{\parallel}$ being built  from two form factors $A_{1,2}$, which would lead to an increase of the
error on $\xi_{\parallel}$ if one takes the errors on $A_{1,2}$ as independent. This problem can actually be avoided because
eq.~(\ref{eq:q20rel}) allows us to extract the error on $\xi_{\parallel}(0)$ from $A_0(0)$\footnote{We increase the error on
$\xi_{\parallel}(0)$ obtained in this way by the small extent to which the LCSR form factors violate the relation
(\ref{eq:q20rel}).
For the form factors from ref.~\cite{Khodjamirian:2010vf} we are still left with determining the error
for the slope-parameter of the $q^2$-dependence of $\xi_{\|}$ from $A_1$ and $A_2$.
Even though the error of $\xi_{\|}$ increases significantly for large $q^2$ because of the unknown correlation of uncertainties in the slopes of $A_1$ and $A_2$,
this has only a minor impact on the errors of optimised observables where the form-factor enters only at next-to-leading order.}.

In ref.~\cite{Jager:2012uw} the authors have chosen a scheme similar to our second scheme, by defining $\xi_{\perp}$ and $\xi_{\parallel}$ in terms of $T_1$ and $A_0$. As discussed above, this explains to some extent the big uncertainties they find in the observable $P_5^{\prime}$. Their scheme differs, however, from our second scheme as they assigned
an ad-hoc $q^2$-dependence to the soft form factors $\xi_{\perp,\parallel}$ which differs from that of the QCD form factors $T_1,A_0$: Equations (\ref{eq:xiT1}) and (\ref{eq:xiA0}) are thus fulfilled only at $q^2=0$. As a consequence, only the $a$-coefficients are correlated in their scheme, and the correlations of $b$- and higher-order coefficients are lost. This reduces the number of correlations artificially from eight, as in our second scheme, to only four in their scheme (if parameters a,b,c are considered). Another difference in their study is that they extract $T_1$ from experimental data on $B\to K^*\gamma$ while we take $T_1$ from its LCSR calculations -- the limits of the first approach compared to our extraction from theoretical computations of the form factors have already been discussed in a  previous section, after eq.~(\ref{eq:xiT1}).
Concerning the extraction of factorisable power corrections, our main differences with respect to the approach in ref.~\cite{Jager:2012uw} are the following: we vary the power correction parameters $a,b,c$ around their (non-zero) fit values with a separate assessment of the uncertainties ($\hat{a}_F-\Delta \hat{a}_F\leq a_F\leq \hat{a}_F\leq \Delta \hat{a}_F$), rather than varying them in the whole range given by the magnitude of the fit value ($-|\hat{a}_F|\leq a_F\leq |\hat{a}_F|$). We have chosen a different renormalization scheme leading to stronger correlations and generally smaller errors.
Furthermore, we do not average different form factor determinations (especially we do not perform average of absolute values, leading to numerical values for the power correction parameters inconsistent with respect to the renormalisation scheme chosen). These differences result in better controlled uncertainties on the angular observables shown in Figure~\ref{fig_SD}.

\section{Non-factorisable power corrections}
\label{sec:NonFacPow}

Even in the situation in which QCD form factors were known exactly, the problem of non-factorisable contributions would persist.
This problem is related to the factorisation of hadronic contributions to $B\to K^*\mu^+\mu^-$ from four-quark and chromo-magnetic
operators where the lepton pair is produced via a virtual photon. At large recoil, the factorisation of the corresponding matrix
elements into form factors, light-cone distribution amplitudes and hard-scattering kernels is a formal prediction of SCET/QCD factorisation at
leading power in the  $1/m_b$ expansion \cite{Beneke:2001at}. At subleading power, however,
new unknown non-perturbative contributions would appear. These power corrections are called non-factorisable,
and appear irrespectively of whether QCD form factors are expressed in terms of soft
form factors or not. An estimate of such power corrections must be included in the predictions.

An approach that has become popular \cite{Egede:2008uy} consists in parametrising both factorisable and non-factorisable
power corrections jointly via a set of complex factors multiplying each transversity amplitude, with typical absolute values
of order $10\%$ (motivated from dimensional arguments) and arbitrary phases\footnote{Shortcomings related to this procedure,
as well as the general problems related to the estimation of power corrections in $B\to K^*\ell\ell$ have been recently reviewed in
Ref.~\cite{Hurth:2013ssa}.}. 
Even if this ad-hoc procedure tends to underestimate the errors associated to individual transversity amplitudes in the vicinity of the zeroes, this is not the case for observables. Error estimates based on this strategy
are expected to give reasonable results for physical observables because they receive contributions from various amplitudes, and left- and right-handed transversity amplitudes do not vanish at the same value of $q^2$ (with the sets of form factors currently available).

In our present analysis we could use the same technique for non-factorisable power corrections alone (since
factorisable power corrections are estimated separately using the more sophisticated methods described in Section \ref{sec:FacPow}), but that would clearly overestimate the effect. Note that the contributions
from electromagnetic and semileptonic operators are free from non-factorisable corrections, so that the terms proportional to $C_{7,9,10}^{(\prime)}$, which are leading contributions, must not be inflated artificially.

Therefore we proceed as follows: in ref.~\cite{Beneke:2001at}, the amplitudes of $\langle K^* \gamma^*|H_{eff}|B\rangle$ are
decomposed in terms of three hadronic form factors ${\cal T}_i(q^2)$, which are re-expressed in terms of Wilson coefficients,
soft form factors, light-cone sum rules and hard-scattering kernels using QCD factorisation. In each of the amplitudes,
we single out the part involving the hadronic form factors ${\cal T}_i^\text{had}$, obtained from the functions ${\cal T}_i$ by
\footnote{The amplitudes ${\cal T}_i$ are defined from $\langle K^* \gamma^*|H_{eff}|B\rangle$ and thus do not
contain contributions proportional to $C_{9,10}^{(\prime)}$. In the presence of right-handed currents
(i.e., chirally-flipped operators $\ord_{i}'$) the set of amplitudes generalizes to ${\cal T}_i^\pm$ (see
e.g., ref.~\cite{Lunghi:2006hc}). Here we use the collective symbol ${\cal T}_i$ for all of them.}
${\cal T}_i^\text{had} = {\cal T}_i|_{C_7^{(\prime)}\to0}$.
Finally, we multiply each of these amplitudes with a complex $q^2$-dependent factor:
\eq{
{\cal T}_i^\text{had}\to \big(1+r_i(q^2)\big) {\cal T}_i^\text{had},
}
with
\eq{r_i(s) = r_i^a e^{i\phi_i^a} + r_i^b e^{i\phi_i^b} (s/m_B^2) + r_i^c e^{i\phi_i^c} (s/m_B^2)^2.}
Let us note at this point that the relationship ${\cal T}_2 = 2E/m_B\,{\cal T}_1$ \cite{Beneke:2001at} does not hold at subleading
power, so that our parameters $r_2$ and $r_3$ for non-factorisable power corrections are unrelated.

We define our central values as the ones with $r_i(q^2)\equiv0$, and estimate the uncertainties from non-factorisable power corrections
by varying $r_i^{a,b,c}\in [0,0.1]$ and $\phi_i^{a,b,c} \in [-\pi,\pi]$ independently, corresponding to a
$\sim 10\%$ correction with an arbitrary phase. The uncertainties for each observable are then obtained by performing a random
scan and taking the maximum deviation from the central values to each side, to obtain (possibly asymmetric) upward and
downward error bars.

\section{Results}
\label{sec:results}

\subsection{SM predictions for angular observables}

\begin{figure}
\centering
\includegraphics[width=6.5cm,height=4.4cm]{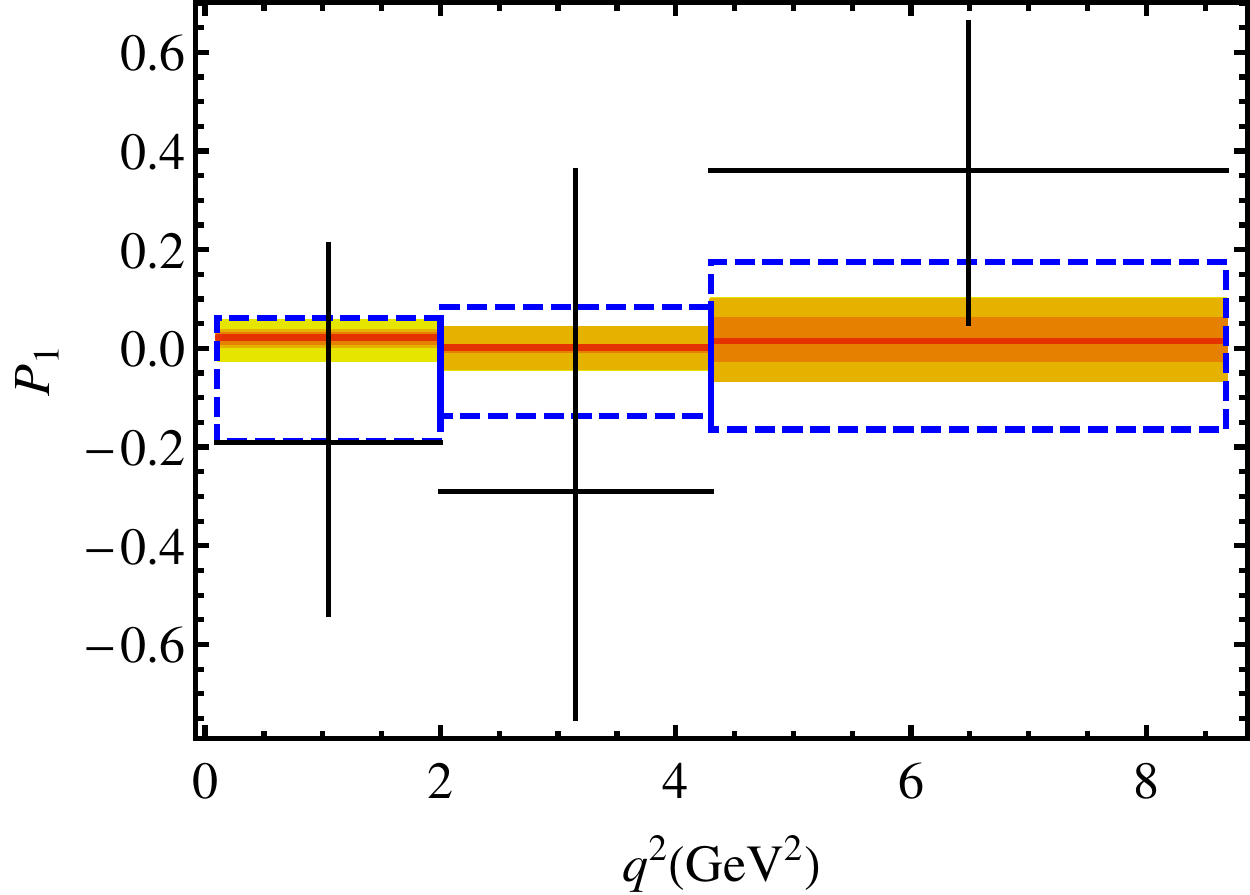}\hspace{1cm}
\includegraphics[width=6.5cm,height=4.4cm]{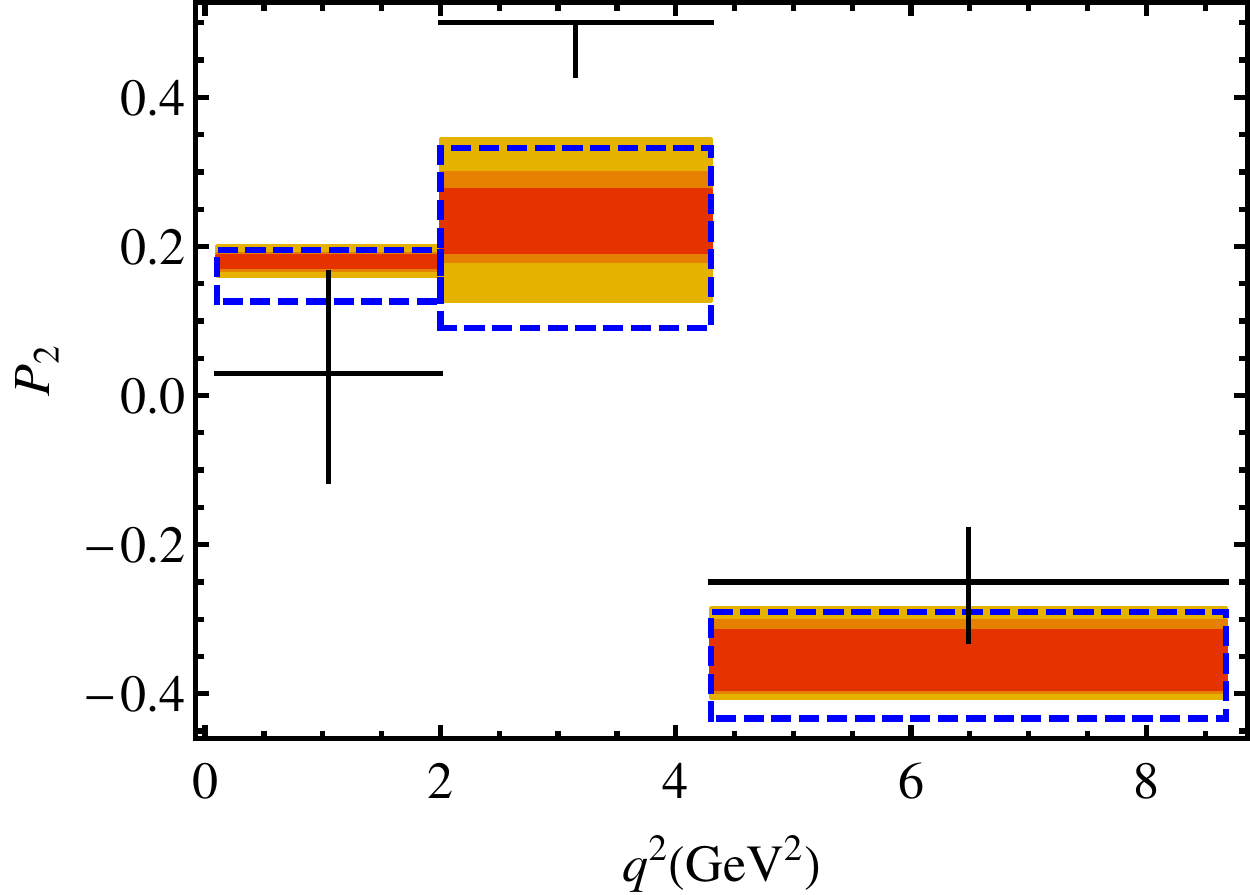}\\[2mm]
\includegraphics[width=6.5cm,height=4.4cm]{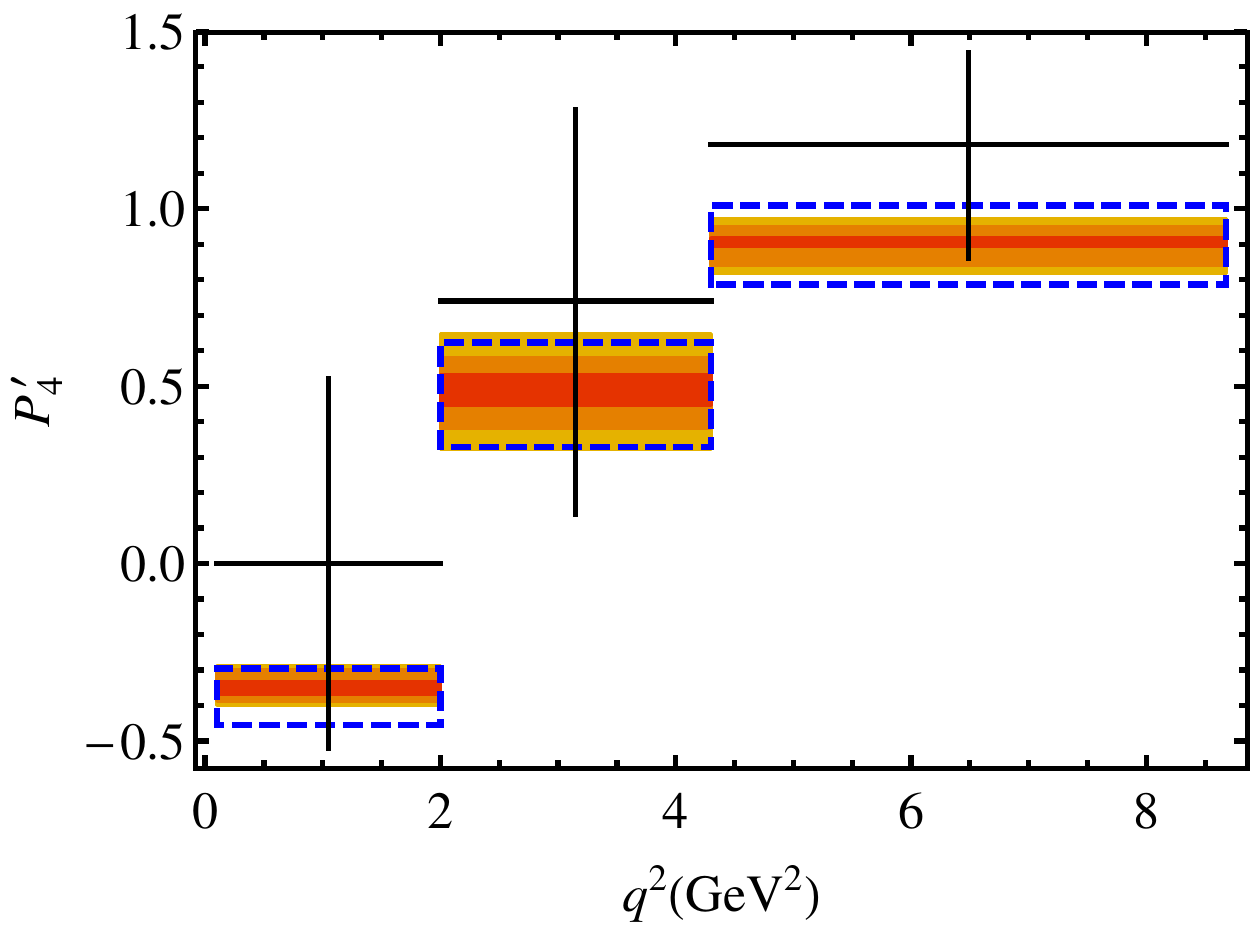}\hspace{1cm}
\includegraphics[width=6.5cm,height=4.4cm]{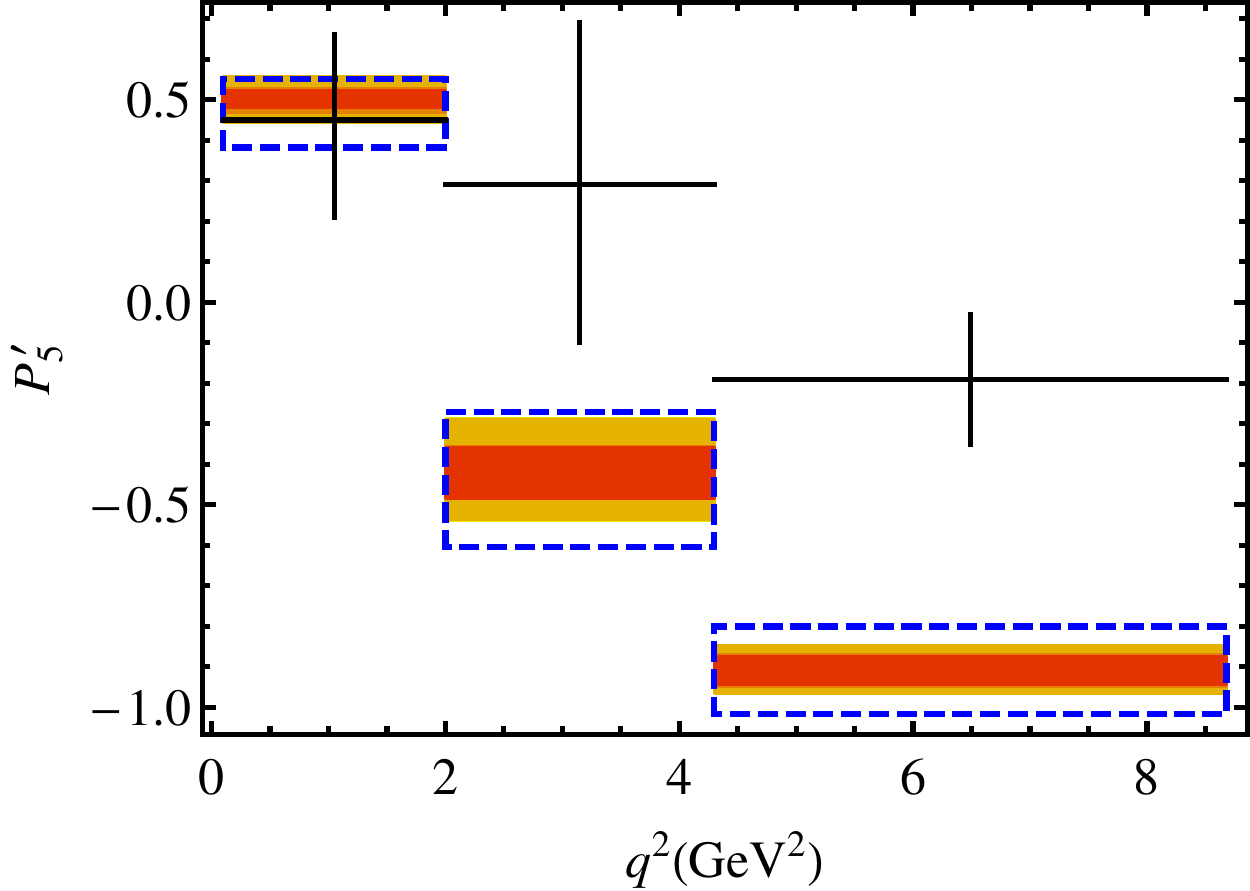}
\caption{SM predictions for the observables $P_1$, $P_2$, $P_4'$, $P_5'$ obtained as described in the text. The bands correspond,
from darker to lighter, to uncertainties from parametric, form factor, factorisable and non-factorisable power corrections,
added sequentially in quadrature. The data
points correspond to experimental data from LHCb~\cite{Aaij:2013iag,Aaij:2013qta}. Blue dashed boxes are predictions in scheme 2.}
\label{fig:res}
\end{figure}

In this section we present the set of SM predictions for the various angular observables. We give results within scheme 1 (where
soft form factors are defined from $V$, $A_1$, $A_2$), which globally leads to smaller uncertainties related to factorisable
power corrections, as detailed in Section~\ref{sec:schemedep}. We do not provide the results for scheme 2 (where
soft form factors are defined from $T_1$, $A_0$), but we do include these in the plots below (Figure~\ref{fig:res}) for comparison. 
We have explored several schemes~\footnote{Besides schemes 1 and 2 discussed in the paper, we have also considered a mixed scheme were soft form factors are defined from $V, A_0$. This scheme leads to very similar results to scheme 2.} and find that
scheme 1 is preferred for many observables. In the case of observables sensitive to $C_9$, of particular interest for the analysis of the deviations observed by LHCb~\cite{Descotes-Genon:2013wba}, an argument in favour of this scheme has been
given in Section~\ref{sec:schemedep}.
We stress that in principle one can choose different schemes for different observables consistently, allowing one to optimise the accuracy of the theory prediction for each individual observables. In global analyses (i.e. global fits), on the other hand, all observables should be calculated using the same scheme because otherwise different observables would depend on different sets of theory parameters $\xi^{(i)}_{\perp,\|}$ and $a^{(i)}_F,b^{(i)}_F,c^{(i)}_F$ and correlations among the predictions for different observables would be lost.

The central value for each observable corresponds to the value obtained by setting all the parameters to their central values,
including factorisable power corrections, as obtained from the central values of the parameters $a_F,b_F,c_F$ in
tables~\ref{tab:fit1} and \ref{tab:fit2}. This is an important difference with respect to previous analyses based on QCD factorisation, where central values correspond
to subleading contributions put to zero. In particular our central values are comparable to those obtained from analyses that
use QCD form factors (e.g., ref.~\cite{Altmannshofer:2008dz}).

Uncertainties related to factorisable and non-factorisable power corrections are computed as described in
Sections~\ref{sec:FacPow} and~\ref{sec:NonFacPow}, and presented separately. The rest of the error analysis is separated into
``parametric" and ``form factors". The first accounts for the variation of all input parameters except form factors
(masses, decay constants, Gegenbauer moments, renormalisation scale, taking the same inputs as in ref.~\cite{Descotes-Genon:2013vna}), and the second for the errors associated to $\xi_{\|,\bot}(q^2)$, inherited from the form factor input in the respective scheme.
For all four types of uncertainties, errors ranges are obtained in the same way, which we
illustrate by focusing on the parametric uncertainties: we make a random flat scan of all relevant parameters (masses, etc.)
simultaneously, within the range given by their ``uncertainty" (error bars given by the PDG~\cite{PDG} in the case of masses, the renormalisation scale between $m_b/2$ and $2m_b$, etc.), while keeping the other sets of parameters (form factors, power corrections) fixed to their
central values. We compute each observable for every point in the scan, and take the corresponding maximum and minimum value. Upward and downward error bars are then obtained by comparing the extreme values with the central values. 

\begin{table}
\small
\centering
\begin{tabular}{@{}lrr@{}}
\hline\hline 
Observable & \begin{minipage}{6cm}\hspace{2cm} KMPW - scheme 1\end{minipage} & \begin{minipage}{6cm}\hspace{2cm} BZ - scheme 1 \end{minipage} \\ 
 \hline 
$\av{P_1}_{[0.1,2]} $  & $0.021_{-0.003}^{+0.004}{}_{-0.010}^{+0.008}{}_{-0.012}^{+0.011}{}_{-0.043}^{+0.034} $ & $0.035_{-0.003}^{+0.005}{}_{-0.000}^{+0.000}{}_{-0.011}^{+0.010}{}_{-0.045}^{+0.035} $ \\ 
 \hline 
$\av{P_1}_{[2,4.3]} $  & $0.000_{-0.002}^{+0.004}{}_{-0.006}^{+0.001}{}_{-0.040}^{+0.040}{}_{-0.013}^{+0.009} $ & $-0.023_{-0.003}^{+0.003}{}_{-0.000}^{+0.000}{}_{-0.057}^{+0.049}{}_{-0.009}^{+0.007} $ \\ 
 \hline 
$\av{P_1}_{[4.3,8.68]} $  & $0.013_{-0.001}^{+0.002}{}_{-0.037}^{+0.046}{}_{-0.069}^{+0.071}{}_{-0.005}^{+0.005} $ & $-0.101_{-0.003}^{+0.002}{}_{-0.000}^{+0.000}{}_{-0.074}^{+0.076}{}_{-0.005}^{+0.005} $ \\ 
 \hline 
$\av{P_1}_{[1,6]} $  & $0.009_{-0.001}^{+0.002}{}_{-0.012}^{+0.009}{}_{-0.040}^{+0.037}{}_{-0.014}^{+0.010} $ & $-0.031_{-0.004}^{+0.003}{}_{-0.000}^{+0.000}{}_{-0.054}^{+0.045}{}_{-0.011}^{+0.009} $ \\ 
 \hline 
$\av{P_1}_{[1,2]} $  & $0.002_{-0.002}^{+0.003}{}_{-0.020}^{+0.015}{}_{-0.023}^{+0.020}{}_{-0.043}^{+0.033} $ & $0.031_{-0.003}^{+0.004}{}_{-0.000}^{+0.000}{}_{-0.019}^{+0.015}{}_{-0.043}^{+0.033} $ \\ 
 \hline 
$\av{P_1}_{[4.3,6]} $  & $0.021_{-0.002}^{+0.004}{}_{-0.033}^{+0.039}{}_{-0.069}^{+0.068}{}_{-0.002}^{+0.002} $ & $-0.071_{-0.002}^{+0.000}{}_{-0.000}^{+0.000}{}_{-0.077}^{+0.077}{}_{-0.003}^{+0.003} $ \\ 
 \hline 
$\av{P_1}_{[6,8]} $  & $0.015_{-0.001}^{+0.003}{}_{-0.039}^{+0.049}{}_{-0.070}^{+0.073}{}_{-0.004}^{+0.004} $ & $-0.104_{-0.003}^{+0.002}{}_{-0.000}^{+0.000}{}_{-0.075}^{+0.077}{}_{-0.005}^{+0.004} $ \\ 
 \hline 
$\av{P_2}_{[0.1,2]} $  & $0.179_{-0.007}^{+0.008}{}_{-0.007}^{+0.006}{}_{-0.015}^{+0.018}{}_{-0.002}^{+0.002} $ & $0.187_{-0.008}^{+0.008}{}_{-0.000}^{+0.000}{}_{-0.016}^{+0.015}{}_{-0.002}^{+0.002} $ \\ 
 \hline 
$\av{P_2}_{[2,4.3]} $  & $0.244_{-0.053}^{+0.030}{}_{-0.038}^{+0.044}{}_{-0.098}^{+0.083}{}_{-0.013}^{+0.010} $ & $0.156_{-0.056}^{+0.035}{}_{-0.000}^{+0.000}{}_{-0.099}^{+0.102}{}_{-0.015}^{+0.011} $ \\ 
 \hline 
$\av{P_2}_{[4.3,8.68]} $  & $-0.344_{-0.050}^{+0.028}{}_{-0.019}^{+0.030}{}_{-0.030}^{+0.041}{}_{-0.003}^{+0.004} $ & $-0.386_{-0.039}^{+0.021}{}_{-0.000}^{+0.000}{}_{-0.022}^{+0.032}{}_{-0.002}^{+0.003} $ \\ 
 \hline 
$\av{P_2}_{[1,6]} $  & $0.106_{-0.054}^{+0.026}{}_{-0.034}^{+0.042}{}_{-0.078}^{+0.071}{}_{-0.010}^{+0.008} $ & $0.034_{-0.052}^{+0.026}{}_{-0.000}^{+0.000}{}_{-0.076}^{+0.082}{}_{-0.011}^{+0.008} $ \\ 
 \hline 
$\av{P_2}_{[1,2]} $  & $0.409_{-0.017}^{+0.017}{}_{-0.016}^{+0.012}{}_{-0.031}^{+0.030}{}_{-0.004}^{+0.004} $ & $0.429_{-0.016}^{+0.015}{}_{-0.000}^{+0.000}{}_{-0.031}^{+0.022}{}_{-0.004}^{+0.004} $ \\ 
 \hline 
$\av{P_2}_{[4.3,6]} $  & $-0.210_{-0.066}^{+0.030}{}_{-0.031}^{+0.044}{}_{-0.057}^{+0.073}{}_{-0.007}^{+0.006} $ & $-0.281_{-0.054}^{+0.023}{}_{-0.000}^{+0.000}{}_{-0.045}^{+0.063}{}_{-0.006}^{+0.005} $ \\ 
 \hline 
$\av{P_2}_{[6,8]} $  & $-0.376_{-0.057}^{+0.026}{}_{-0.016}^{+0.025}{}_{-0.024}^{+0.034}{}_{-0.002}^{+0.003} $ & $-0.412_{-0.043}^{+0.019}{}_{-0.000}^{+0.000}{}_{-0.016}^{+0.025}{}_{-0.002}^{+0.003} $ \\ 
 \hline 
$\av{P'_4}_{[0.1,2]} $  & $-0.352_{-0.016}^{+0.019}{}_{-0.031}^{+0.047}{}_{-0.031}^{+0.039}{}_{-0.009}^{+0.009} $ & $-0.316_{-0.017}^{+0.024}{}_{-0.001}^{+0.001}{}_{-0.034}^{+0.042}{}_{-0.010}^{+0.010} $ \\ 
 \hline 
$\av{P'_4}_{[2,4.3]} $  & $0.485_{-0.039}^{+0.047}{}_{-0.094}^{+0.082}{}_{-0.125}^{+0.129}{}_{-0.009}^{+0.010} $ & $0.628_{-0.036}^{+0.041}{}_{-0.002}^{+0.001}{}_{-0.131}^{+0.112}{}_{-0.009}^{+0.010} $ \\ 
 \hline 
$\av{P'_4}_{[4.3,8.68]} $  & $0.902_{-0.008}^{+0.014}{}_{-0.060}^{+0.045}{}_{-0.056}^{+0.050}{}_{-0.004}^{+0.005} $ & $0.993_{-0.005}^{+0.010}{}_{-0.000}^{+0.000}{}_{-0.049}^{+0.043}{}_{-0.003}^{+0.004} $ \\ 
 \hline 
$\av{P'_4}_{[1,6]} $  & $0.476_{-0.034}^{+0.041}{}_{-0.095}^{+0.091}{}_{-0.111}^{+0.116}{}_{-0.008}^{+0.009} $ & $0.594_{-0.031}^{+0.037}{}_{-0.002}^{+0.002}{}_{-0.117}^{+0.103}{}_{-0.008}^{+0.009} $ \\ 
 \hline 
$\av{P'_4}_{[1,2]} $  & $-0.186_{-0.023}^{+0.034}{}_{-0.053}^{+0.059}{}_{-0.069}^{+0.091}{}_{-0.011}^{+0.011} $ & $-0.105_{-0.028}^{+0.044}{}_{-0.002}^{+0.002}{}_{-0.080}^{+0.095}{}_{-0.012}^{+0.012} $ \\ 
 \hline 
$\av{P'_4}_{[4.3,6]} $  & $0.842_{-0.015}^{+0.018}{}_{-0.069}^{+0.052}{}_{-0.076}^{+0.067}{}_{-0.004}^{+0.004} $ & $0.950_{-0.010}^{+0.012}{}_{-0.000}^{+0.000}{}_{-0.067}^{+0.054}{}_{-0.003}^{+0.003} $ \\ 
 \hline 
$\av{P'_4}_{[6,8]} $  & $0.930_{-0.011}^{+0.012}{}_{-0.053}^{+0.038}{}_{-0.052}^{+0.046}{}_{-0.004}^{+0.005} $ & $1.019_{-0.009}^{+0.008}{}_{-0.000}^{+0.000}{}_{-0.044}^{+0.040}{}_{-0.003}^{+0.004} $ \\ 
 \hline 
$\av{P'_5}_{[0.1,2]} $  & $0.505_{-0.024}^{+0.015}{}_{-0.028}^{+0.014}{}_{-0.049}^{+0.045}{}_{-0.012}^{+0.011} $ & $0.506_{-0.025}^{+0.016}{}_{-0.000}^{+0.000}{}_{-0.048}^{+0.042}{}_{-0.013}^{+0.012} $ \\ 
 \hline 
$\av{P'_5}_{[2,4.3]} $  & $-0.411_{-0.072}^{+0.050}{}_{-0.015}^{+0.017}{}_{-0.101}^{+0.109}{}_{-0.020}^{+0.016} $ & $-0.436_{-0.068}^{+0.048}{}_{-0.000}^{+0.000}{}_{-0.097}^{+0.095}{}_{-0.019}^{+0.016} $ \\ 
 \hline 
$\av{P'_5}_{[4.3,8.68]} $  & $-0.902_{-0.043}^{+0.025}{}_{-0.021}^{+0.019}{}_{-0.041}^{+0.043}{}_{-0.006}^{+0.006} $ & $-0.853_{-0.036}^{+0.021}{}_{-0.000}^{+0.000}{}_{-0.047}^{+0.048}{}_{-0.006}^{+0.006} $ \\ 
 \hline 
$\av{P'_5}_{[1,6]} $  & $-0.412_{-0.070}^{+0.042}{}_{-0.045}^{+0.026}{}_{-0.089}^{+0.096}{}_{-0.017}^{+0.014} $ & $-0.416_{-0.064}^{+0.039}{}_{-0.000}^{+0.000}{}_{-0.086}^{+0.083}{}_{-0.017}^{+0.014} $ \\ 
 \hline 
$\av{P'_5}_{[1,2]} $  & $0.331_{-0.045}^{+0.029}{}_{-0.006}^{+0.013}{}_{-0.081}^{+0.074}{}_{-0.017}^{+0.015} $ & $0.315_{-0.048}^{+0.032}{}_{-0.001}^{+0.001}{}_{-0.084}^{+0.073}{}_{-0.018}^{+0.016} $ \\ 
 \hline 
$\av{P'_5}_{[4.3,6]} $  & $-0.832_{-0.060}^{+0.027}{}_{-0.013}^{+0.018}{}_{-0.057}^{+0.058}{}_{-0.007}^{+0.007} $ & $-0.802_{-0.052}^{+0.024}{}_{-0.000}^{+0.000}{}_{-0.059}^{+0.059}{}_{-0.007}^{+0.006} $ \\ 
 \hline 
$\av{P'_5}_{[6,8]} $  & $-0.934_{-0.047}^{+0.024}{}_{-0.022}^{+0.021}{}_{-0.038}^{+0.039}{}_{-0.005}^{+0.005} $ & $-0.880_{-0.039}^{+0.020}{}_{-0.000}^{+0.000}{}_{-0.044}^{+0.045}{}_{-0.005}^{+0.005} $ \\ 
 \hline 
\hline
\end{tabular}
\caption{SM predictions for the observables $P_1$, $P_2$, $P_4'$, $P_5'$ in various bins, computed in scheme 1, where
the soft form factors are determined from ($V$,$A_1$,$A_2$). First error is parametric, second
is form factors, third is factorisable power corrections and fourth is non-factorisable power corrections. The first column
(KMPW) is obtained with LCSR input from ref.~\cite{Khodjamirian:2010vf} and the second one (BZ) from ref.~\cite{Ball:2004ye}. Slight differences in the central values with respect to refs.~\cite{Descotes-Genon:2013wba,Descotes-Genon:2013vna} are due to a different numerical value for the charm pole mass, which we take here as $m_c=1.47\pm 0.20$ GeV.}
\label{tab:res1}
\end{table}

Table~\ref{tab:res1} contains our results for a selected set of observables in scheme 1, where we give both results based on
form-factor input from \cite{Khodjamirian:2010vf} (KMPW) and \cite{Ball:2004ye} (BZ). The corresponding predictions for other
observables are collected for reference in Appendix~\ref{app2}. We note that for optimised observables and for input taken
from KMPW, parametric uncertainties, form factor uncertainties and uncertainties from factorisable power corrections are
usually of the same order of magnitude, while uncertainties from non-factorisable power corrections are typically smaller.
For ''non-optimised observables`` uncertainties are dominated by the form factor input as expected. For input taken from BZ,
the uncertainties stemming from the form factors are generally smaller, in particular they are completely negligible for
optimised observables.
In Figure~\ref{fig:res} we illustrate the predictions corresponding to table~\ref{tab:res1} and in this case to
KMPW form factors together with the experimental data points from the LHCb analyses of in refs.~\cite{Aaij:2013iag,Aaij:2013qta}.
In these figures we add sequentially and quadratically the four different sets of uncertainties as in Table~\ref{tab:res1}.
We include also the predictions in scheme 2 for comparison, noting that
they generally lead to larger uncertainties in $P_1$, $P_2$, $P_4'$, $P_5'$.

\subsection{Impact of $c\bar{c}$ loops}

Our computation includes contributions from $c\bar{c}$ loops, through factorisable contributions as well as 
non-factorisable contributions with hard-gluon exchanges.
As already mentioned in the introduction, the size of the remaining long-distance contribution from $c\bar{c}$ loops is a debated issue, with some contributions considered in ref.~\cite{Khodjamirian:2010vf} for $B\to K^*\mu\mu$ and further work (unfortunately only for $B\to K\mu\mu$) in ref.~\cite{khodjimannel}. We have not considered these contributions up to now explicitly, even though they are partly encoded in the power corrections discussed in the previous sections. Indeed, these contributions do not stand on the same footing as the
factorisable power corrections discussed in Sec.~\ref{sec:FacPow}. While we presented a systematic procedure to estimate the size in the latter case, here we can only rely on a partial computation existing in the literature \cite{Khodjamirian:2010vf}. 

In ref.~\cite{Khodjamirian:2010vf} the soft-gluon contribution originating from the insertion of 4-quark operators ${\cal O}_{1,2}^c$ and penguin operators ${\cal O}_{3-6}$ induces a positive contribution inside $C_9^{\eff}$.
For an overall estimate of non-perturbative contributions from hadronic operators, we take the terms $\Delta C_9$ in Ref~\cite{Khodjamirian:2010vf}, which include the LO perturbative contribution from $O_{1,2}$ together with non-factorisable soft-gluon emission from the charm loop. 
In order to separate the long-distance contribution,
we subtract the perturbative contribution from  $\Delta C_9$ (using
eq.~(7.14) and table 2 of ref.~\cite{Khodjamirian:2010vf}), to obtain the (three) functions $\tilde g(q^2)$ according to eq.~(5.6) of ref.~\cite{Khodjamirian:2010vf}. The results should match well the functions $\tilde g$, at least below $4$~GeV$^2$, computed for $m_c=1.05$ GeV. In order to gauge the possible shift in our central values (computed at the reference value $m_c=1.47$ GeV), we shift $\tilde g$ by $-35\%$ as indicated in Table 1 of ref.~\cite{Khodjamirian:2010vf}. This gives three ranges of variations (one for each function $\tilde g$), from which we construct a single band using the following parametrization \cite{Descotes-Genon:2013wba}:
\begin{equation}
\delta C_9^\text{LD}(q^2) = \frac{a+bq^2(c-q^2)}{q^2(c-q^2)}
\end{equation}
with $a\in [2,7]$ GeV$^4$, $b\in [0.1,0.2]$ and $c\in [9.3,9.9]$ GeV$^2$. The resulting band contains all three $\tilde g$ functions (and their errors) in the range $1<q^2<9$ GeV$^2$. We add this contribution to each amplitude ${\cal A}_i^{L,R}$ by substituting:
\begin{equation}
\C{9} \to \C{9} + s_i \delta C_9^\text{LD}(q^2)\ .
\end{equation}
The parameters $s_i$ are varied independently in the range $[-1,1]$ so that: (i) the contributions to different amplitudes are not artificially correlated, (ii) the possibility of long-distance contribution with opposite signs in the different amplitudes is considered. We emphasize that this method might be overestimating the effect due to (ii) (only one sign corresponds to the computation in ref.~\cite{Khodjamirian:2010vf}, the other is only considered here to remain conservative and is not supported by the results of this reference). We also note that the perturbative charm-loop contributions are already included in our predictions up to NLO, while the effects discussed here are the soft-gluon contributions and the non-perturbative extrapolation to $q^2>4$~GeV$^2$.

\begin{table}
\small
\centering
\begin{tabular}{@{}crrrrrrr@{}}
\hline\hline 
Observable & [0.1,2] &[2,4.3] &[4.3,8.68] &[1,6] &[1,2] &[4.3,6] & [6,8]  \\ 
 \hline 
$\av{P_1} $  & ${}_{-0.091}^{+0.067} $ & ${}_{-0.051}^{+0.041} $ & ${}_{-0.061}^{+0.088} $ & ${}_{-0.031}^{+0.026} $ & ${}_{-0.120}^{+0.089} $ & ${}_{-0.028}^{+0.041} $ & ${}_{-0.061}^{+0.087} $ \\ 
 \hline 
$\av{P_2} $  & ${}_{-0.003}^{+0.004} $ & ${}_{-0.065}^{+0.052} $ & ${}_{-0.048}^{+0.057} $ & ${}_{-0.060}^{+0.052} $ & ${}_{-0.011}^{+0.011} $ & ${}_{-0.063}^{+0.064} $ & ${}_{-0.042}^{+0.051} $ \\ 
 \hline 
$\av{P'_4} $  & ${}_{-0.185}^{+0.237} $ & ${}_{-0.092}^{+0.095} $ & ${}_{-0.089}^{+0.057} $ & ${}_{-0.087}^{+0.091} $ & ${}_{-0.108}^{+0.118} $ & ${}_{-0.076}^{+0.064} $ & ${}_{-0.085}^{+0.055} $ \\ 
 \hline 
$\av{P'_5} $  & ${}_{-0.133}^{+0.093} $ & ${}_{-0.114}^{+0.098} $ & ${}_{-0.082}^{+0.062} $ & ${}_{-0.102}^{+0.088} $ & ${}_{-0.125}^{+0.090} $ & ${}_{-0.079}^{+0.066} $ & ${}_{-0.078}^{+0.058} $ \\ 
 \hline 
$\av{P_3} $  & ${}_{-0.003}^{+0.004} $ & ${}_{-0.006}^{+0.009} $ & ${}_{-0.006}^{+0.008} $ & ${}_{-0.005}^{+0.007} $ & ${}_{-0.005}^{+0.007} $ & ${}_{-0.004}^{+0.006} $ & ${}_{-0.005}^{+0.006} $ \\ 
 \hline 
$\av{P'_6} $  & ${}_{-0.011}^{+0.010} $ & ${}_{-0.005}^{+0.005} $ & ${}_{-0.005}^{+0.005} $ & ${}_{-0.004}^{+0.005} $ & ${}_{-0.011}^{+0.008} $ & ${}_{-0.003}^{+0.003} $ & ${}_{-0.004}^{+0.004} $ \\ 
 \hline 
$\av{P'_8} $  & ${}_{-0.018}^{+0.016} $ & ${}_{-0.005}^{+0.005} $ & ${}_{-0.004}^{+0.003} $ & ${}_{-0.005}^{+0.005} $ & ${}_{-0.009}^{+0.009} $ & ${}_{-0.003}^{+0.002} $ & ${}_{-0.003}^{+0.003} $ \\ 
 \hline 
$\av{A_\text{FB}} $  & ${}_{-0.010}^{+0.017} $ & ${}_{-0.020}^{+0.021} $ & ${}_{-0.041}^{+0.051} $ & ${}_{-0.021}^{+0.023} $ & ${}_{-0.020}^{+0.022} $ & ${}_{-0.030}^{+0.034} $ & ${}_{-0.042}^{+0.050} $ \\ 
 \hline 
$\av{F_L} $  & ${}_{-0.044}^{+0.062} $ & ${}_{-0.019}^{+0.018} $ & ${}_{-0.045}^{+0.037} $ & ${}_{-0.021}^{+0.021} $ & ${}_{-0.039}^{+0.039} $ & ${}_{-0.026}^{+0.024} $ & ${}_{-0.044}^{+0.037} $ \\ 
 \hline 
$\av{S_3} $  & ${}_{-0.021}^{+0.015} $ & ${}_{-0.005}^{+0.004} $ & ${}_{-0.010}^{+0.014} $ & ${}_{-0.004}^{+0.003} $ & ${}_{-0.018}^{+0.014} $ & ${}_{-0.003}^{+0.005} $ & ${}_{-0.010}^{+0.015} $ \\ 
 \hline 
$\av{S_4} $  & ${}_{-0.038}^{+0.049} $ & ${}_{-0.019}^{+0.021} $ & ${}_{-0.021}^{+0.017} $ & ${}_{-0.019}^{+0.022} $ & ${}_{-0.022}^{+0.026} $ & ${}_{-0.017}^{+0.017} $ & ${}_{-0.020}^{+0.016} $ \\ 
 \hline 
$\av{S_5} $  & ${}_{-0.053}^{+0.032} $ & ${}_{-0.039}^{+0.035} $ & ${}_{-0.037}^{+0.032} $ & ${}_{-0.037}^{+0.034} $ & ${}_{-0.060}^{+0.046} $ & ${}_{-0.033}^{+0.028} $ & ${}_{-0.036}^{+0.031} $ \\ 
 \hline 
$\av{S_{6s}} $  & ${}_{-0.022}^{+0.013} $ & ${}_{-0.028}^{+0.027} $ & ${}_{-0.068}^{+0.055} $ & ${}_{-0.031}^{+0.029} $ & ${}_{-0.030}^{+0.026} $ & ${}_{-0.046}^{+0.040} $ & ${}_{-0.067}^{+0.056} $ \\ 
 \hline 
\hline 
\end{tabular}
\caption{Estimates for the errors in binned observables arising from long-distance charm-loop effects,
as described in the text.\label{tab:res4}}
\end{table}

\begin{figure}
\centering
\includegraphics[width=6.5cm,height=4.2cm]{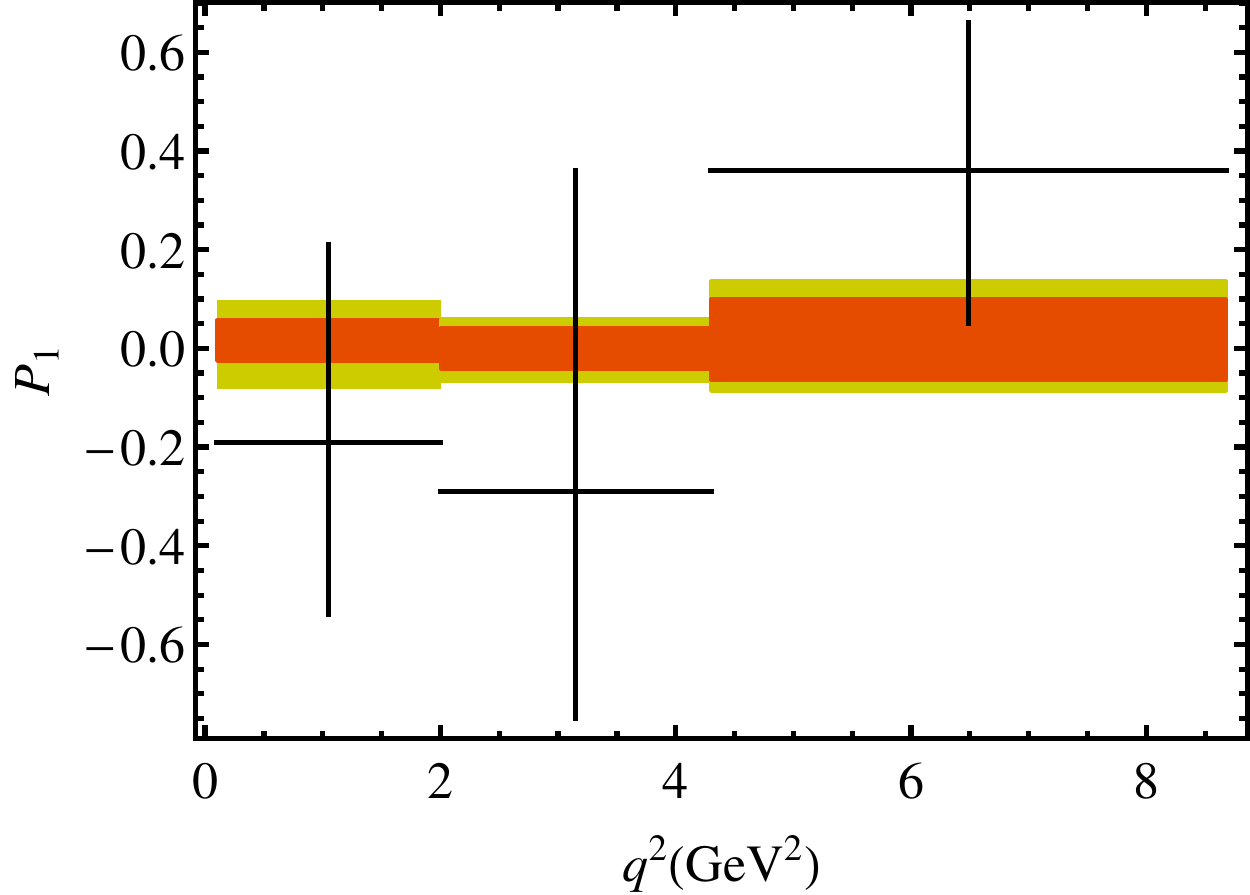}\hspace{1cm}
\includegraphics[width=6.5cm,height=4.2cm]{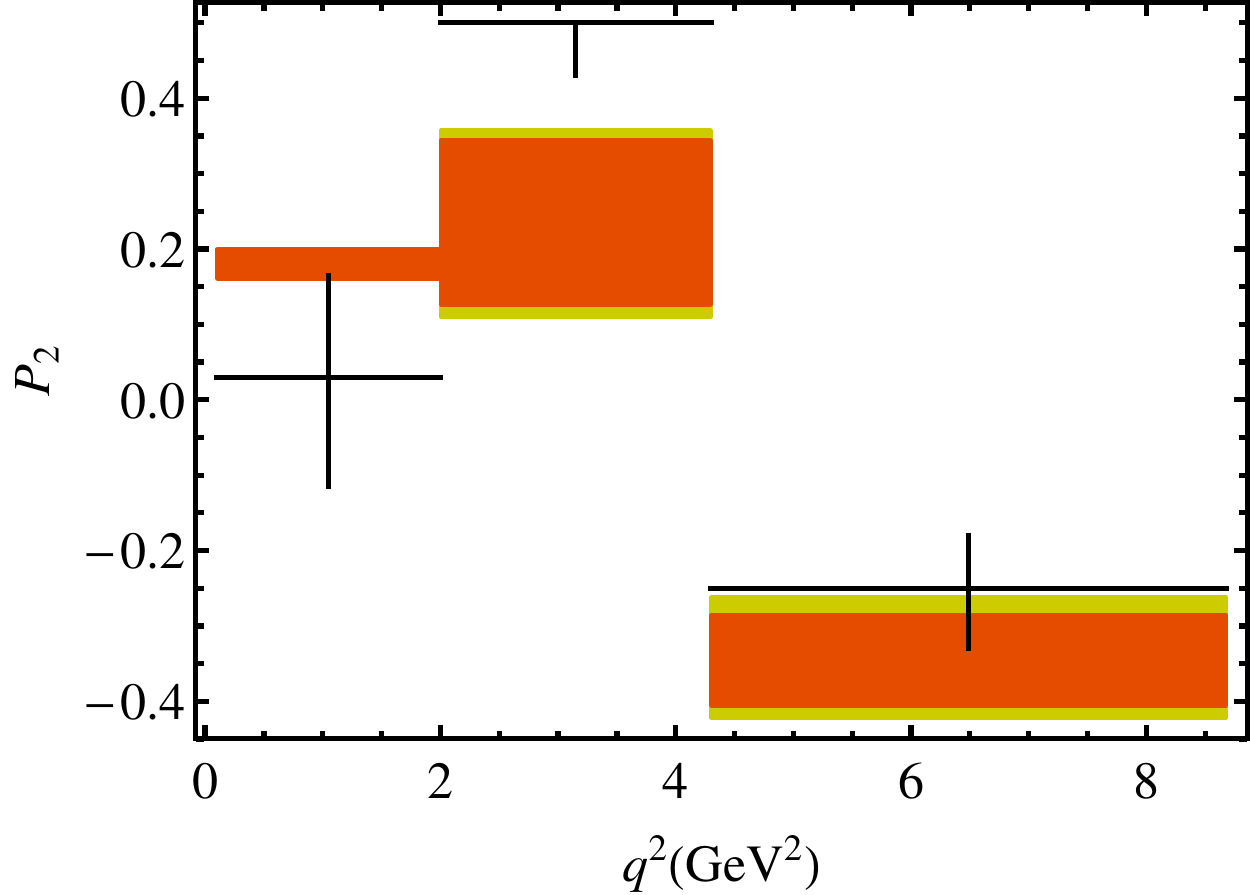}\\[2mm]
\includegraphics[width=6.5cm,height=4.2cm]{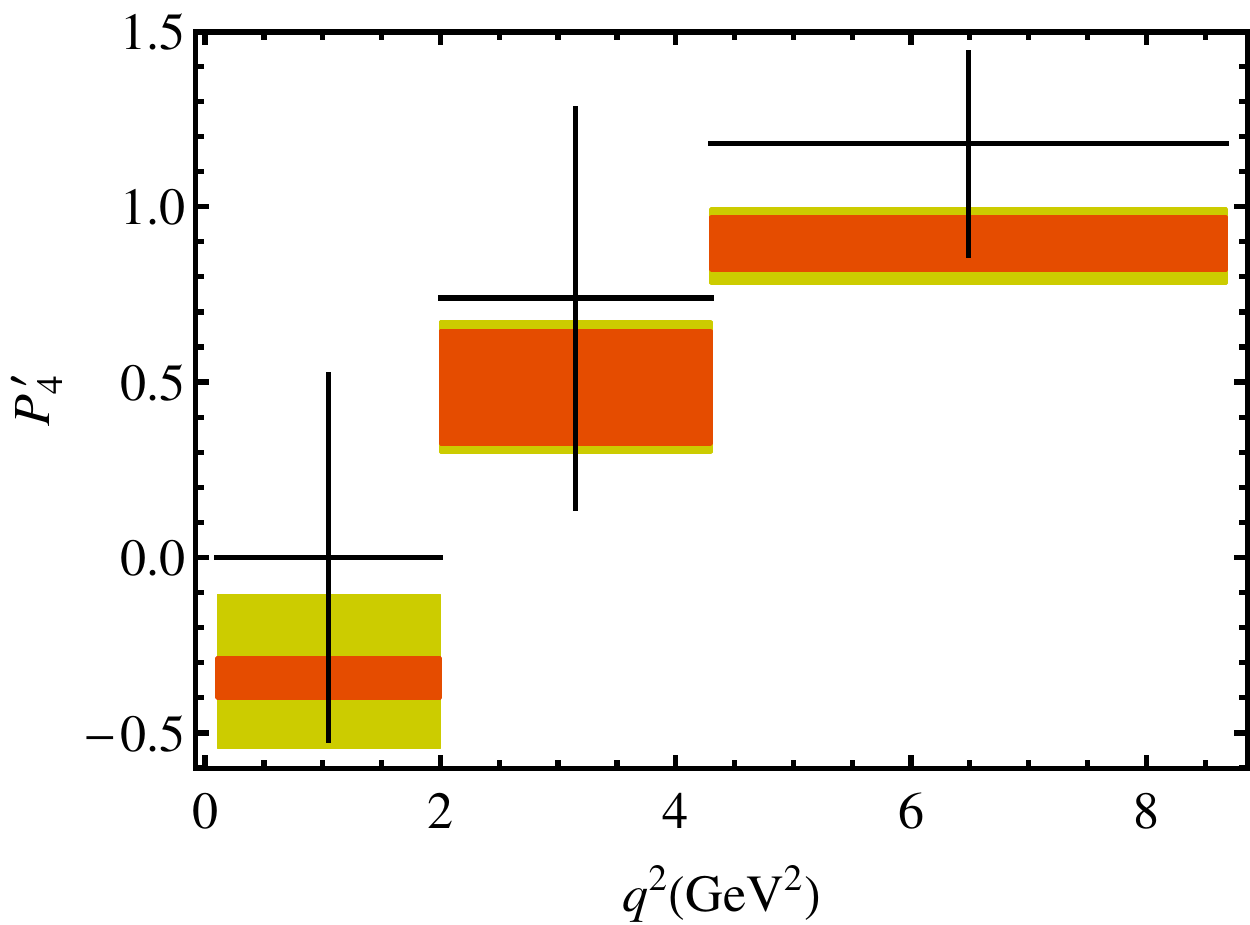}\hspace{1cm}
\includegraphics[width=6.5cm,height=4.2cm]{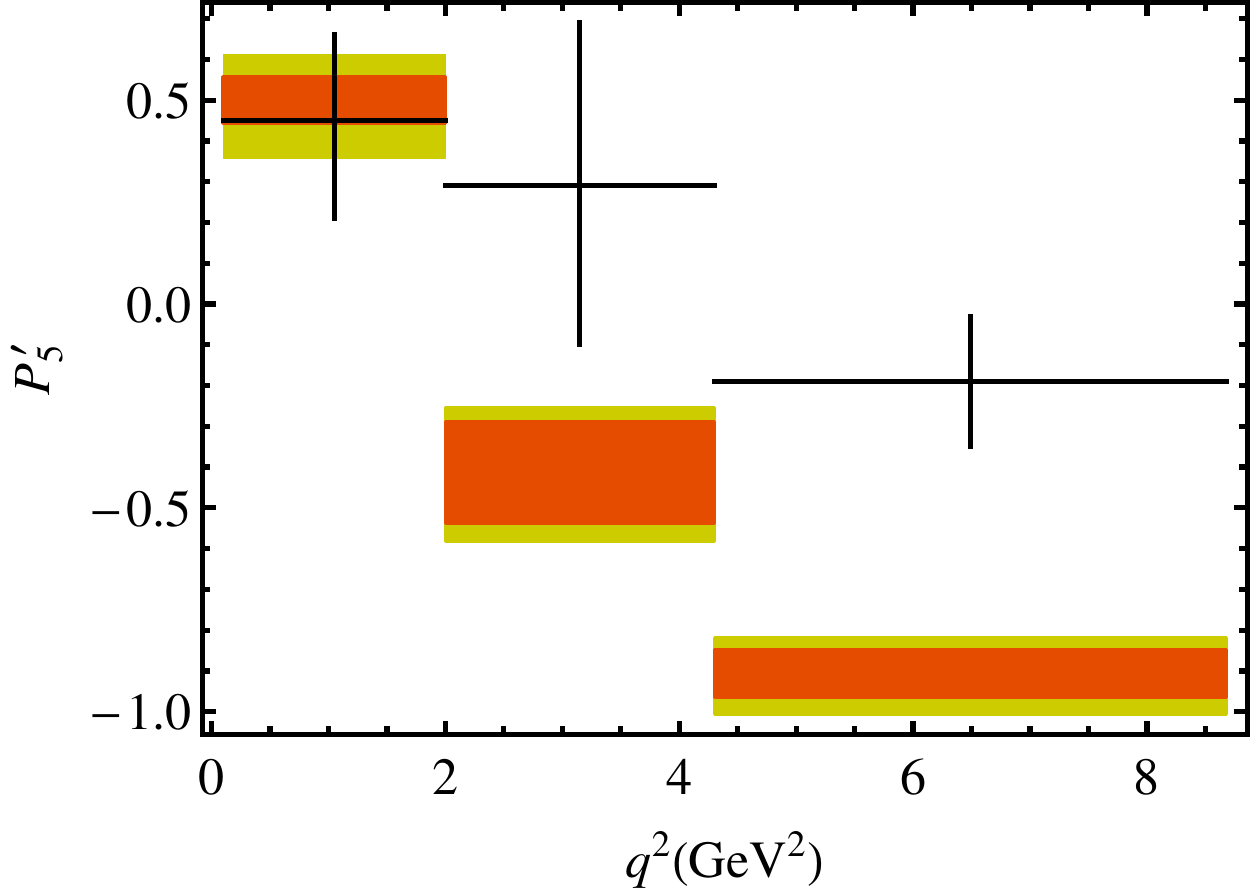}
\caption{SM predictions for the observables $P_1$, $P_2$, $P_4'$, $P_5'$ obtained as described in the text. The bands correspond
to all uncertainties added in quadrature, not including (dark) and including (light) our
 estimate of long-distance charm-loop effects. The data
points correspond to experimental data from LHCb~\cite{Aaij:2013iag,Aaij:2013qta}.
\label{fig:res2}}
\end{figure}

All binned observables are then computed, fixing all parameters to their central values, except for $a,b,c$ and
$s_{\perp,\|,0}$, which are varied within the given ranges. We perform a random scan over these parameters and obtain maximum and minimum values for each observable. Comparing these values to the results with $s_i=0$ (which
correspond to the central values of our predictions in Table~\ref{tab:res1}) we obtain the positive and negative error bars
collected in Table~\ref{tab:res4}. This procedure will be called approach A  in the following. Table~\ref{tab:res4} summarizes our estimates of these effects.  We also show our results
in Figure~\ref{fig:res2} where the long-distance $c\bar{c}$ correction is displayed as a separate band. These plots constitute our
predictions including charm-loop effects.

In order to be conservative in estimating these error we have also followed another approach B, where we evaluate all
relevant observables and check on the difference between the central values obtained including and not including the
long-distance contribution described in ref.~\cite{Khodjamirian:2010vf}. In order to do this comparison, we took the charm
contribution at the same order and also at the indicated reference mass $m_c(2 m_c)=1.05$ GeV as in
ref.~\cite{Khodjamirian:2010vf}. The result of this comparison gives us an estimate of the size of the associated error. Assuming a linear dependence on $m_c$ for the normalisation of the functions $\tilde{g}(1\ {\rm GeV}^2)$ as given in Table 1 in ref.~\cite{Khodjamirian:2010vf}, and translating this into a linear dependence of $\Delta C_9(1\ {\rm GeV}^2)$ in eq.~(7.14) of the same reference, we have also studied the impact of varying $m_c$ from 1.05 GeV up to 1.5 GeV. We found that this approach yields an uncertainty substantially smaller than the default approach A outlined above.

Several comments are in order in relation with Figure~\ref{fig:res2} and Table~\ref{tab:res4}. First, it appears that the impact of the long-distance contribution remains
small up to 8 GeV$^2$ (even a little bit above 8 GeV$^2$ the effect is not very significant). Secondly,
 even if the computation done in ref.~\cite{Khodjamirian:2010vf} implies a definite sign for $\delta C_9^{c\bar{c},{\rm LD}}$,
 following approach A the errors are enlarged to cover the values corresponding to the opposite sign,
 as we interpret the $\delta C_9^{c\bar{c},\rm{LD}}$ extracted from ref.~\cite{Khodjamirian:2010vf} as the expected size of long-distance charm-loop effects rather than taking it at face value.
  As a third remark, we find that
 for some observables the slope is more important for the induced uncertainties than the proximity to the
 resonance region. In other words, an observable with a steep slope like $P_5^\prime$ has a larger error in the intermediate
 region (from 2 to 6 GeV$^2$), due to the significant uncertainty on its slope in this region,  than in the plateau
 (from 6 to 8 GeV$^2$) where the uncertainty is limited. In the case of $P_5^\prime$ in the [6-8] bin, the contribution computed in ref.~\cite{Khodjamirian:2010vf}  would tend to enhance the anomaly: however, there is not much space left below the
 SM prediction without long-distance charm contributions, so that the impact of the latter is small. A similar situation
 occurs in $P_4^\prime$ but not for $P_2$.
 
Other approaches to estimate the uncertainties due to $c\bar{c}$ long-distance contributions could have been followed, but in our opinion, they could yield misleading estimates. First, the original calculation done in
ref.~\cite{Khodjamirian:2010vf} re-expresses these long-distance effects, entering in all observables via $C_9^{\eff}$ as done here. Other approaches~(e.g. ref.~\cite{Jager:2012uw}) choose to transfer the long-distance effect to $C_7$. Even though this can always be done in principle, one should be careful to distinguish an estimate of the impact of $c\bar{c}$-loop contributions on $C_7$ from that on actual observables, as the latter have generally different (and bin-dependent) sensitivities to $C_7$. Including a specific estimate for charm-loop corrections in $C_7$ might thus overestimate the uncertainties induced in observables in some energy ranges and underestimate in others.

A second comment concerns the symmetrisation of errors. In the above procedure, we have split $\Delta C_9$ into the contributions from short and long distances in the $c\bar{c}$ contributions as given ref.~\cite{Khodjamirian:2010vf}. Our errors are obtained varying the sign of the long-distance contribution only (the short-distance part being known from perturbation theory).  We would have obtained artificially enhanced uncertainties, if we had varied the sign of the whole $\Delta C_9$ contribution, which would have corresponded to a ``wrong'' sign for the perturbative contribution.

\section{Conclusions}
\label{sec:Conclu}
The rare $B\to K^*\mu^+\mu^-$ decay has been under recent scrutiny after the LHCb experiment reported deviations with respect to the Standard Model in several observables at large $K^*$ recoil. Following an analysis based on 
QCD factorisation,
these observables $P_i^{(\prime)}$ have been designed to be less sensitive to hadronic uncertainties than the angular coefficients of the differential decay rate.
The central issue consists in separating soft contributions ${\cal O}(\Lambda)$ from hard contributions ${\cal O}(m_b)$ in the expressions of the $B\to K^*$ hadronic form factors and subsequently the helicity amplitudes involved in $B\to K^*\mu^+\mu^-$ decay, leading to a cancellation of the soft form factors in suitable ratios of
angular observables. This cancellation is however only valid at the leading order in QCD factorisation, and the sensitivity to hadronic form factors re-enters through subleading corrections, either hard (computable as a series in $\alpha_s$) or soft (estimated on dimensional grounds as $\Lambda/m_b$).

It was recently claimed that the latter corrections, also known as power corrections, 
could yield much larger uncertainties than expected for the observables measured at LHCb. This would naturally decrease the sensitivity of these observables to New Physics and reduce the significance of the observed deviations. We have reassessed this claim by adopting and improving the analysis strategy described in ref.~\cite{Jager:2012uw} to extract the size of the factorisable power corrections, related to the re-expression of the QCD form factors in terms of soft form factors. We consider sets of QCD form factors obtained from light-cone sum rules, identify two soft form factors and compute the power corrections by taking the difference between QCD form factors and their representation as the sum of soft form factors and perturbative corrections. After the QCD form factors are split in their various constituents (soft form factors, perturbative corrections and power corrections), we have shown how to compute observables making the maximal use of the information on the non-zero central values, the uncertainties and the correlations of the power corrections.

It has been demonstrated that in the analysis of factorisable power corrections, the choice of a renormalisation scheme to define the soft form factors out of the QCD form factors has an important impact on the results. Choosing an inappropriate renormalisation scheme, generating large, weakly correlated power corrections for the most relevant form factors for the observables of interest, might lead the factorisable power corrections to induce abnormally large errors for the $B\to K^*\mu^+\mu^-$ angular observables.
We have shown that another (well-documented) scheme yields significantly smaller power corrections than the one chosen in ref.~\cite{Jager:2012uw} for two different sets of QCD form factors, corresponding to the most recent determinations of QCD form factors based on light-cone sum rules. We have computed angular observables within this renormalisation scheme, keeping track of the correlations between the various parameters describing the factorisable power corrections. The results for angular observables are collected in Table~\ref{tab:res1} (with a decomposition into the various sources of uncertainties) and illustrated in Figure~\ref{fig:res} (where results for two different schemes are presented, including non-factorisable corrections).  We have also discussed the (small) impact of long-distance contributions from charm loops based on estimates available in the literature, as seen in Table~\ref{tab:res4} and Figure~\ref{fig:res2}.

In the process of writing this paper, other issues have been raised concerning the role played by long-distance $c\bar{c}$ loops both for $B\to K\mu^+\mu^-$ and $B\to K^*\mu \mu$, which are discussed in the appendix of this paper. Both discussions (on the size of power corrections and on the impact of long-distance charmonium dynamics) are useful to reduce the uncertainties attached to the predictions for $B\to K^*\mu^+\mu^-$ angular observables. A thorough check of the uncertainties attached to these observables is essential to assess the anomaly currently observed in LHCb data, and ultimately confirm its connection with New Physics.

\section{Acknowledgements}

We thank Alexander Khodjamirian, Ulrick Egede, Thorsten Feldmann, Dirk Seidel, Nicola Serra, David Straub, Danny van Dyk, Roman Zwicky
and Tobias Hurth for useful discussions.
J.V. is funded by the Deutsche Forschungsgemeinschaft (DFG) within research unit FOR 1873 (QFET).
J.M. and L.H. acknowledge support from FPA2011-25948, SGR2009-00894.


\appendix

\section{Can charm loops always mimic New Physics ?}
\label{app1}

In a recent article~\cite{zwicky}, it has been claimed that  the observed anomaly in $P_5^\prime$ could be solved thanks to a 350\% correction with respect to the factorisation approximation coming from charm-resonance effects.\footnote{ For simplicity, we call "charm-resonance effects"  a contribution  from  charm loops following ref.\cite{zwicky}, irrespectively  of its origin (long-distance QCD and/or  new $\bar{b} s c\bar c$ structures). The term "New Physics" will be used exclusively  to refer to a new high-scale contribution to one of the Wilson coefficients $C_i^{(\prime)}$ $i=7,9,10$.}
We will not discuss here all the implicit and model-dependent assumptions involved in this approach and
necessary to relate $e^+ e^-$ data with the dynamics of charmonia in $B \to K\mu^+\mu^-$, if the resonance model chosen (with a sum of Breit-Wigner charmonia) can be extrapolated far from the resonance peaks, whether two constant ``fudge factors'' are enough to capture all the departures of $B \to K\mu^+\mu^-$  data from this specific resonance model in both low- and large-$K$ recoil regimes, or if the same fudge factors hold unchanged for $B \to K\mu^+\mu^-$ and $B \to K^*\mu^+\mu^-$. Instead we will take the solution proposed in ref.~\cite{zwicky}, which implies a very specific $q^2$-dependent form for long-distance $c\bar{c}$ contributions and propose three different tests that can be implemented to assess the validity of this proposal.

A  fit to all $b\to s\ell\ell$ observables including this model of contribution for long-distance $c\bar{c}$ loops could shed some light on the global coherence of such a proposal, even though a likely outcome of this fit would be a slight improvement compared to the standard modelisation, as it would include two new free parameters. In this situation, more specific observables could provide a  more clear-cut test of this model for charm-loop contributions.
We start defining the semileptonic  coefficients as in ref.~\cite{zwicky}
\bea \label{wc1} C_9^{\eff}=C_9 + a_{fac} \eta_c h_c(q^2) + h_{rest}(q^2), \quad  \quad \quad  \label{wc2} C_9^{\prime\eff}=C_9^\prime + a_{fac} \eta_c^\prime h_c(q^2),\eea  
where $a_{fac} \sim 0.6$ stems from factorisation, $\eta_c$ and $\eta_c^\prime$ are pre-factor parameters found to be large and negative from the analysis of low-recoil $B\to K\mu^+\mu^-$ differential branching ratio (whereas the standard expectations would be $\eta_c=1,\eta'_c=0$). The function $h_c(q^2)$ describes long- and short-distance from charm loops, through a dispersive relation applied to a Breit-Wigner model for the observed $c\bar{c}$ resonances in $\sigma(e^+e^-\to$ hadrons). $h_{rest}(q^2)$ stands for the sub-leading contributions from other flavours which are very tiny and will be neglected for the rest of the discussion.

It was argued in ref.~\cite{zwicky} that the result of fitting the data at low recoil for ${\cal B}(B^+ \to K^+ \mu^+\mu^-)$, sensitive to the sum $C_9^+=C_9^{\eff}+C_9^{\prime \eff}$, imposes
 \beq \label{eta} \eta_c+\eta_c^\prime\sim-2.5 \eeq
It is important to remark at this point that ref.~\cite{zwicky} assumes implicitly that the Standard Model holds in order to obtain eq.~(\ref{eta}) by  combining $e^+ e^-$ data  with ${\cal B}(B^+ \to K^+ \mu^+\mu^-)$ data. Indeed, if New Physics affected $B \to K \mu^+\mu^-$ data, the sum eq.~(\ref{eta}) could be reduced substantially, so that charm-resonance effect could not accommodate the  $P_5^\prime$ anomaly contrary to what is stated in ref.~\cite{zwicky}.

Our approach here is to explore patterns in designed observables that cannot be explained in the Standard Model by the modification of the prefactors $\eta_c$ and $\eta_c^\prime$ to the charm-loop contribution entering $C_9^{\eff}$ and $C_9^{\prime\eff}$ as proposed in ref. \cite{zwicky}. Such patterns would thus require New Physics even if the charm loop model of ref.~\cite{zwicky} is valid.  In the following, we will mostly work  under the hypothesis of no New Physics
  \beq C_{7,9,10}^{NP}=0, \quad \quad \quad C^\prime_{7,9,10}=0. \eeq Within this framework  $C_9^{\prime\eff}$ would contain only  $a_{fac}\eta_c^\prime h_c (q^2)$ but not New Physics. However, in some cases we will relax this hypothesis and allow for New Physics to illustrate how certain conditions change.

The authors of ref.~\cite{zwicky} find in agreement with ref.~\cite{Descotes-Genon:2013wba} that in order to explain the $B\to K^*\mu^+\mu^-$ anomaly, a scenario is favoured where  the (effective) Wilson coefficients $C_9^\eff$ and $C_{9}^{\prime\eff}$ receive new contributions with $\Delta C_9^{\eff} \simeq \Delta C_{9}^{\prime\eff}$. { They claim} that these new contributions could be generated from resonant charm loops rather than from high-scale new physics. In this appendix we discuss three tests on the forthcoming data which could { disentangle} the two proposals.
 The first test will consist in identifying an observable for which the charm-loop contribution eqs.~(\ref{wc1})-(\ref{eta}) alone cannot mimic the contribution from a New Physics source. The second test is based on observing the presence of these charm contributions in the related $b \to d$ transition decay $B \to \pi \mu^+\mu^-$.  Finally, the third more qualitative test aims at disentangling the effect of two large negative parameters $\eta_c,\eta'_c$ from a true New Physics contribution to $C_9$ and $C_9^\prime$.

\subsection{Test 1: $P_1$ strikes back}

Our first test will focus on $B\to K^*\mu^+\mu^-$ angular observables for which specific values cannot be accommodated by the charm-loop model eqs.~(\ref{wc1})-(\ref{eta}) but are allowed in New Physics models. If Eq.(\ref{eta}) holds, the largest impact of this charm-loop model should be expected in observables sensitive to $C_9^+$. 
In the context of the $B\to K^*\mu^+\mu^-$ decay  the inspection of the transversity amplitudes (see \cite{Matias:2012xw}) suggests that an observable proportional to $A_\perp^{L,R}$ will do the job, such as

\beq \label{q} Q_{[6\leq q^2 \leq 8]}(q^2)={1+P_1(q^2)}=\frac{2 |A_{\perp}|^2}{|A_{\perp}|^2+|A_\||^2} \eeq where it is understood that $|A_i|^2=|A_i^L|^2+|A_i^R|^2$ is the sum of the corresponding left and right transversity amplitudes and the subscript indicates the relevant range for $q^2$. 
{ From this observable $Q$ one immediately obtains two more observables
  $$R=Q \times F_T=F_T+ 2 S_3,  \quad   \quad \quad  S=R \times d\Gamma/dq^2.$$ 
  
 As a probe of the Wilson coefficients, $S$ plays in $B \to K^* \mu^+ \mu^-$ a similar role to ${\cal B}(B \to K \mu^+ \mu^-)$. Both are only a function of 
  $C_9^+= C_9^{\eff}+ C_9^{\prime \eff}$, contrary to ${\cal B}(B \to K^*\mu^+ \mu^-)$ which is a function of $C_9^+$ and $C_9^-= C_9^{\eff}- C_9^{\prime \eff}$.}
    
  
 LHCb~\cite{Aaij:2013qta} found for the wide third bin \beq \label{data} Q_{[4.3 \leq q^2 \leq 8.68]} \sim 1.36 \pm 0.30\eeq  There is also a previous measurement by CDF~\cite{cdf1,cdf2} but with a very large uncertainty.

One can understand the discriminating power of this observable with the following argument~\footnote{We will not consider the impact of power corrections here. In any case, we have seen that at most they tend to shift up $P_1$ approximately by +0.11 in this bin, which can be expected to be the maximum value above zero reached by this bin within the SM. A scan over $\eta_c$ and $\eta_c^\prime$ satisfying eq.~(\ref{eta})  confirms this expectation.}.
In the SM within the large-recoil range but for $q^2$ not small (between 6 to 8 GeV$^2$), the electromagnetic piece of the amplitude proportional to $C_7$ is subleading and the semileptonic contributions linked to $C_{9,10}$ dominate. At leading order one can approximate this observable in this region as
\beq \label{q2} 
Q_{[6\leq q^2 \leq 8]}(q^2) \sim  \frac{|C_9^{+}|^2 + |C_{10}|^2}{|C_9^{+}|^2/2 + |C_9^{-}|^2/2+|C_{10}|^2} 
\eeq
 In the standard case $\eta_c=1, \eta_c^\prime=0$ which implies $C_9^+=C_9^-$
and   $Q_{[6 \leq q^2 \leq 8]}\sim 1$ (in agreement with our SM prediction of $P_1^{\rm[6,8]}=0.015^{+0.088}_{-0.080}$).

Under the hypothesis that future data will significantly increase the significance of the deviation from one of eq.~(\ref{data}) we will explore the implication of the condition $Q_{[6\leq q^2\leq 8]} > 1$, which translates using eq.~(\ref{q2}) into the constraint
\beq \label{cond} {\rm Re} C_9^{\eff} C_9^{\prime\eff *} > 0 \quad \Rightarrow \quad  (C_9+ a_{fac} \eta_c {\rm Re} h_c) (a_{fac} \eta_c^\prime {\rm Re} h_c) > 0\eeq
where $h_c$ is real  in this region according to ref.~\cite{zwicky}. This equation requires the same sign for $C_9^{\eff}$ and $C_9^{\prime\eff}$, which implies two solutions for $\eta_c$, $\eta_c^\prime$:

\begin{itemize}

\item[I.] $\eta_c^\prime<0$ (both coefficients negative):  then using eq.~(\ref{eta}) and eq.~(\ref{cond})  one finds $$-2.5 < \eta_c < {\rm Max}[- C_9/(a_{fac} {\rm Re}h_c)]_{[6\leq q^2 \leq 8]}=-2.6$$ 
This condition is obviously impossible to fulfil. The right-hand side term reaches its maximum 
at $q^2=$8 GeV$^2$ defining the most favourable situation, still impossible to satisfy.
This is not surprising because the sign of $C_9^{\eff}$ can be changed only for very large negative $\eta_c$. Notice that even if at first sight a NP contribution of the type  $C_9^{NP}<0$ could extend the allowed range and allow this solution, one should first 
reassess the determination of $\eta_c+\eta_c^\prime$ which was performed in the SM, and second, check that this value of $\eta_c$
allows for a zero in $A_{FB}$ (see Test 3 below).

\item[II.] $\eta_c^\prime>0$ (both coefficients positive): then using the same equations one gets
$$-2.5 > \eta_c > {\rm Min}[- C_9/(a_{fac} {\rm Re} h_c)]_{[6\leq q^2 \leq 8]}=-4.3$$
This range of values for $\eta_c$ is also excluded because for these values of $\eta_c$, $P_2$ (or $A_{FB}$) has no zero (see Test 3). %

\end{itemize}

The power of this test can be illustrated by the cases considered in ref.~\cite{zwicky}. One of the illustrative examples ($\eta_c=0, \eta_c^\prime=-2.5$) in ref.~\cite{zwicky} yields $Q\sim 0.5$ in  the bin [4.3,8.68]~\footnote{$P_1$ and consequently $Q$ can be inferred from the values of $P_2$ and $P_{4,5}^\prime$ in \cite{nicola} (see also Eq.(\ref{P2P4P5})) or determined by direct computation.} which is disfavoured by LHCb measurements, 
 and another one ($\eta_c=-2.5, \eta_c^\prime=0$) is also disfavoured due to the lack of zero in $A_{FB}$ ({or $P_2$}) {(see Test 3 and Figure~12 in ref.~\cite{zwicky})}.

In summary, if an accurate measurement of the last bin of $P_1$  (bin [6,8]) shows 
a clear preference for $Q_{[6,8]}>1$, it cannot be accommodated by the solution $\eta_c+\eta_c^\prime \sim -2.5$ {\it with no New Physics contributions}.

A value of $Q_{[6,8]}$ exceeding its SM prediction $Q\sim 1.11$ can be attained in the presence of certain NP, for example in the presence right-handed currents.
  The subleading terms in $Q_{[6,8]}$ can become important when NP is present: for instance,  if $C_7^\prime \sim 0.06$, $C_{10}^\prime \sim -1$ and $C_9^{NP}\sim -1$ (allowed at 2$\sigma$ according to ref.~\cite{Descotes-Genon:2013wba})  a large deviation of order $Q \sim 1.4$ is generated while keeping $\eta_c=1$ and $\eta_c^\prime=0$. Notice that if NP is also switched on, a solution with $\eta_c\neq 1$ and $\eta_c^\prime \neq 0$ is allowed. This test (if $Q>1$) provides an explicit example where eqs.~(\ref{wc1})-(\ref{eta}) alone would fail in giving an explanation, unless New Physics is allowed. In this sense this first test should be understood more as a test on the presence of New Physics generating $Q>1$ that cannot be polluted by charm loop than a test of eqs.~(\ref{wc1})-(\ref{eta}) themselves. In the case where $Q \sim 1$  the test loses its discriminating power.

Finally, let us recall that $P_1$ is constrained by $P_4^\prime$ by  $P_1 \leq 1-P_4^{\prime 2}$ \cite{nicola}. A measurement of the [6,8] bin of $P_4^\prime$ constrains $Q_{[6\leq q^2 \leq 8]}\leq 2-P_{4 \, [6\leq q^2\leq 8]}^{\prime 2}$ (up to small corrections due to binning).

\subsection{Test 2: $B^+ \to \pi^+ \mu^+\mu^-$}

This test relies on the similarities and differences between  $B^+\to K^+ \mu^+\mu^-$ and  $B^+ \to \pi^+ \mu^+\mu^-$ decays. 
Since  $B^+\to K^+ \mu^+\mu^-$ is a $b \to s$ transition while $B^+ \to \pi^+ \mu^+\mu^-$ comes from $b \to d$, New Physics will affect them differently: in certain models one could expect to see a deviation in the $b\to s$ transition and no deviation in the corresponding $b \to d$ decay.  Under these circumstances,
  the large impact of the charm loop model eqs.~(\ref{wc1})-(\ref{eta}) should affect both decays and could be  
 tested directly. One would expect to see  the same pattern in ${\cal B}(B^+ \to K^+ \mu^+\mu^-)$ and in ${\cal B}(B^+ \to \pi^+ \mu^+\mu^-)$ in the low-$q^2$ region ($1 \leq q^2 \leq   8$ GeV$^2$), namely values below the SM prediction due to the large destructive charm-loop interference.

One should however take care of the different CKM structure involved in the two decays.
The charm loop has the CKM coefficient $V_{cb}V_{cD}^*=-V_{tb}V_{tD}^*(1+V_{ub}V_{uD}^*/V_{tb}V_{tD}^*)$ (with $D=d,s$). Whereas the second term is doubly Cabibbo-suppressed for $D=s$, it remains Cabibbo-allowed for $D=d$ and should be included in the discussion, as shown in eq.(16) of ref.~\cite{ali}. When moving from $b\to s$ to $b\to d$
 the coefficient in front of the charm loop inside $C_9^{\eff}$  becomes
\beq h(m_c,q^2) \to \left(1- \frac{R_b}{R_t} e^{i \alpha}\right) h(m_c,q^2) \eeq
Taking $R_b/R_t \sim 0.4$ and $\alpha \sim 90^\circ$ the real part of the coefficient remains positive and dominates. Thus, following ref.~\cite{zwicky}
and substituting $h_c$ by $\eta_c h_c$ (with $\eta_c$ a large negative parameter), one would expect to see a suppression of ${\cal B}(B^+ \to \pi^+ \mu^+\mu^-)$ with respect to the SM prediction with $\eta_c=1$, $\eta_c^\prime=0$. Indeed, as can be seen from ref.~\cite{ali}, this branching ratio involves $|C_9^+|^2$, and an illustrative back-of-the-envelope computation indicates that for $q^2$=8 GeV$^2$, one has $C_9^+\sim 0.2+i 1.6$ and $|C_9^+|^2 \sim 2.5$ for $\eta_c+\eta'_c=-2.5$, whereas  $C_9^+\sim 5.6-i 0.6 $ and $|C_9^+|^2 \sim 32.1$ for $\eta_c+\eta'_c=1$, confirming the expected suppression of ${\cal B}(B^+ \to \pi^+ \mu^+\mu^-)$ with respect to the SM prediction with $\eta_c=1$, $\eta_c^\prime=0$. { Also one should take into account when comparing those modes the possible impact of annihilation contributions (see, for instance, \cite{zwickyiso} for $B \to K$ case).} 

In summary, a measurement of ${\cal B}(B^+ \to \pi^+ \mu^+\mu^-)$ in the low-$q^2$ region ($1 \leq q^2 \leq 8$ GeV$^2$) above  the SM or in perfect agreement with SM  would disfavour the charm-loop destructive effect eqs.~(\ref{wc1})-(\ref{eta}). On the contrary if data in this region is below the SM prediction as in  ${\cal B}(B^+ \to K^+ \mu^+\mu^-)$, one cannot disentangle between a charm loop effect or a New Physics effect affecting also the $b\to d$   transition. 
The present situation is that there is a first measurement done by LHCb \cite{lhcbpi}    in the entire range of $q^2$ 
$${\cal B}(B^+ \to \pi^+ \mu^+\mu^-)=(2.3 \pm 0.6 {\rm(stat.)} \pm 0.1 {\rm(syst.)}) \times 10^{-8}$$ 
and two compatible SM theory predictions ${\cal B}(B^+ \to \pi^+ \mu^+\mu^-)=(1.88^{+0.32}_{-0.21}) \times 10^{-8}$ \cite{ali} and  ${\cal B}(B^+ \to \pi^+ \mu^+\mu^-)=(2.0 \pm 0.2) \times 10^{-8}$\cite{alternative}. Even if this comparison  would seem to be already now  in conflict with the model in ref.~\cite{zwicky}, we insist that the comparison must be done only in the low-$q^2$ region, where the discussion is much simpler due to the absence of resonances. According to ref.~\cite{ali} the SM prediction is, with $\eta_c=1$, $\eta_c^\prime=0$, 
$$ {\cal B}(B^+ \to \pi^+ \mu^+\mu^-)_{[1 \leq q^2 \leq 8]}=(0.58 ^{+0.09}_{-0.06}) \times 10^{-8}$$
If LHCb measures this bin with a measurement above or in agreement with this value,  the charm-loop model eqs.~(\ref{wc1})-(\ref{eta}) would need to be revised.

\subsection{Test 3: Zeroes and branching ratio}

This third category of tests will be focused on identifying observables able to disentangle the large contributions from the long-distance charm-loop model eqs.~(\ref{wc1})-(\ref{eta}) from a New Physics contribution to the short-distance Wilson coefficients.
We will  focus first on the zero/zeroes of the observable $P_2$  and consider later the behaviour of  the branching ratio of $B \to K \mu^+\mu^-$ at the upper end of the large-recoil region.
We should remark that this last category of tests  is extremely challenging  experimentally.

An independent constraint on $\eta_c$ in the SM comes from the existence of a zero in $P_2$ (or $A_{FB}$)~\footnote{LHCb \cite{Aaij:2013iag} found a zero in $A_{FB}$ at $q_0^2=4.9\pm 0.9$ GeV$^2$.}. At leading order,  $\eta_c$ must fulfil the equation
\beq \label{zero} -2 m_b M_B C_7^{\eff} \frac{1}{s_i} = C_9 + a_{fac} \eta_c {\rm Re} h_c(s_i) \eeq
where $s_i$ stands for the zero(es) in $q^2$.
We can impose that there must exist a zero at leading order (at NLO the position of the zero is typically shifted by $\sim 1$ GeV$^2$) between, say, 2 and 6 GeV$^2$  (a smaller allowed range in $q^2$ implies a stronger constraint on $\eta_c$). Using our inputs and the variation of $h_c$ in this range, we find that \beq 
\label{bound} \eta_c \gtrsim -2
\eeq
Combining this bound with the solution  eq.~(\ref{eta}) advocated in ref.~\cite{zwicky}, we see that $\eta_c^\prime$ cannot  vanish. If New Physics is allowed only in $C_9$ and $C_9^\prime$ (but not in $C_{10}^\prime$), Eq.({\ref{zero}) is unchanged but the bound becomes more constraining in the case of a negative New Physics contribution to $C_9$, reducing substantially the impact of the charm loop on $C_9^{\eff}$:
\beq \label{bound2}
\eta_c \gtrsim -2 - C_9^{NP}/(a_{fac} {\rm Re} h_c(s_i))
\eeq
   Using Appendix B of ref.~\cite{Matias:2012xw} one can easily generalise this expression to NP affecting other Wilson coefficients.

Eq.~(\ref{zero}) also shows that for a  subset of {\it negative} values for $\eta_c$ fulfilling the  bound eq.~(\ref{bound}), a second zero in $P_2$ would arise at a higher value of $q^2$ still within the large-recoil region. Notice that there is no second zero if $\eta_c=1$, with or without New Physics. The observation of a second zero {\it below} 8 GeV$^2$ would give a strong hint in favour of the charm-loop model eqs.~(\ref{wc1})-(\ref{eta}). Conversely, not finding this second zero does not disprove directly this model, but it would push $\eta_c^\prime$ towards large negative values implying a large negative $C_9^{\prime\eff}$ that has to be tested against other observables. However, checking if such a  second zero exists so close to 8 GeV$^2$ seems very challenging from the experimental point of view.

Further comments are in order concerning how a value of $\eta_c\neq 1$ would affect various observables at the upper end of the large-recoil region. One can also see that $P_2$ and $P_5^\prime$ should vanish at the $J/\psi$ peak -- and the speed at which they tend to zero is related to $h_c$. The reason is that in the numerator of these observables there is a cancellation of the quadratic term in $C_9^{\eff}$ (see Appendix B in \cite{Matias:2012xw})  which implies that the numerator is at most linear in the function $h_c(s)$. This cancellation does not occur in the denominator that contains terms proportional to $h_c(s)^2$. If $\eta_c=1$ the divergent behaviour  of $h_c(s)$ is  not visible until $q^2>8.5$ GeV$^2$ but for large and negative $\eta_c$ the effect of the divergence is enhanced and the tendency to zero should be more evident before 8 GeV$^2$. Let us stress that this vanishing behaviour is different from the second zero of $P_2$ discussed in the previous paragraph.

As a side remark it is interesting to notice that the zeroes of  $P_2$ are related to the sign of $P_5^\prime$. One can show easily using the relation~\cite{nicola}
\beq \label{P2P4P5}
P_2 = \frac{1}{2} \left[ P_4^\prime P_5^\prime + \sqrt{(-1+P_1+P_4^{\prime 2})(-1-P_1+P_5^{\prime 2})} \right]
\eeq
that  at the point where $P_2=0$, eq.~(\ref{P2P4P5}) requires $P_5^\prime$  to be negative (given that $P_4^\prime>0$ in agreement with data), which implies by continuity that the curve of $P_5^\prime$ is below $P_2$ in the vicinity of the points where $P_2=0$. This should happen independently of the value of $\eta_c$ and at each zero. Interestingly, this might have implications on the relative positions
of $P_2$ and $P_5^\prime$ in the bins near the zero(es) of $P_2$.

Finally, an important difference between refs.~\cite{zwicky} and \cite{Descotes-Genon:2013wba} comes from the $q^2$-dependence of the Wilson coefficients. In the charm-loop model eqs.~(\ref{wc1})-(\ref{eta}), $C_9^{+}$ decreases with $q^2$. The same occurs for $C_9^{\eff}$ if $\eta_c$ is negative. In ref.~\cite{Descotes-Genon:2013wba} where $\eta_c=1$ and $\eta_c^\prime=0$, both  $C_9^+$ and $C_9^{\eff}$ increase with $q^2$.  If $\eta_c+\eta_c^\prime$ turns out to be large and negative, this should be seen in observables sensitive to $C_9^+$: for instance ${\cal B}(B \to K \mu^+\mu^-)$ { or $S$} should exhibit a more pronounced suppression from [4.3,6] to [6,8] than expected from a standard calculation using $\eta_c=1$ (see Figure~10 in ref.~\cite{khodjimannel}).

In summary the tests proposed in this section aim at disentangling a New Physics  contribution to $C_9$ from a charm loop effect. They rely  on the behaviour induced by the charm-loop model in ref.~\cite{zwicky} in angular observables at the upper end of the large-recoil region, where the sensitivity to a large negative $\eta_c$ parameter (if any) should be more visible. The required accuracy to perform such tests exceeds what can be achieved experimentally for the moment, but presents very interesting challenges for the future.

\newpage

\section{Factorisable $\alpha_s$ corrections}
\label{app3}

In this appendix we collect the expressions for the factorisable $\alpha_s$ corrections $\Delta F^{\alpha_s}$ appearing in the soft form factor representation eq.~(\ref{eq:FF}). They can be found in ref.~\cite{Beneke:2000wa} where they have been derived for a renormalization scheme defining $\xi_{\perp}$ and $\xi_{\|}$ in terms of $V$ and $A_0$. Translating their results to our preferred scheme with $(\xi^{(1)}_{\perp},\xi^{(1)}_{\perp})$ defined from $(V,A_1,A_2)$ we obtain 
\begin{eqnarray}
   \Delta V^{(1)\alpha_s}&=&0,\nonumber\\
   \Delta A_1^{(1)\alpha_s}&=&\Delta A_2^{(1)\alpha_s}\,=\,\mathcal{O}(\alpha_s^2),\nonumber\\
   \Delta A_0^{(1)\alpha_s}&=&\frac{E(q^2)}{m_{K^*}}\,\xi^{(1)}_{\|}(q^2)\,\left(\frac{1}{\Delta}-1\right),\nonumber\\
   \Delta T_1^{(1)\alpha_s}&=&C_F\alpha_s(\mu_b)\,\xi^{(1)}_{\perp}(q^2)\,\left[\log\frac{m_b^2}{\mu_b^2}-L\right]\,+\,
                           C_F\alpha_s(\mu_b)\,\delta T_1,\\
   \Delta T_2^{(1)\alpha_s}&=&C_F\alpha_s(\mu_b)\,\frac{2E(q^2)}{m_B}\,\xi^{(1)}_{\perp}(q^2)\,\left[\log\frac{m_b^2}{\mu_b^2}-L\right]\,+\,C_F\alpha_s(\mu_h)\,\delta T_2,\nonumber\\
   \Delta T_3^{(1)\alpha_s}&=&C_F\alpha_s(\mu_b)\left(\xi^{(1)}_{\perp}(q^2)\,\left[\log\frac{m_b^2}{\mu_b^2}-L\right]\,-\,
                           \xi^{(1)}_{\|}(q^2)\,\left[\log\frac{m_b^2}{\mu_b^2}+2L\right]\right) +\,C_F\alpha_s(\mu_h)\,\delta T_3\,, \nonumber
\end{eqnarray}
where $L=-(2E/(m_B-2E))\log(2E/m_B)$, $\mu_b$ and $\mu_h$ are typical scales for hard processes and $\Delta$ is defined in eq.~(66) of \cite{Beneke:2001at}. The spectator scattering terms are given by
\begin{equation}
  \delta T_1\,=\,\frac{m_B}{4E}\Delta F_{\perp},\hspace{1cm} \delta T_2=\frac{1}{2}\Delta F_{\perp},\hspace{1cm}
  \delta T_3\,=\,\delta T_1\,+\,2\frac{m_{K^*}}{m_B}\,\left(\frac{m_B}{2E}\right)^2\,\Delta F_{\|},
\end{equation}
with $\Delta F_{\perp,\|}$ defined in eq.~(59) of \cite{Beneke:2000wa}. In the scheme with $(\xi^{(2)}_{\perp},\xi^{(2)}_{\perp})$ defined from $(T_1,A_0)$ we get
\begin{eqnarray}
   \Delta V^{(2)\alpha_s}&=&-\frac{m_B+m_{K^*}}{m_B}\,\left\{C_F\alpha_s(\mu_b)\,\xi^{(2)}_{\perp}\,\left[\log\frac{m_b^2}{\mu_b^2}-L\right]\,+\,C_F\alpha_s(\mu_h)\,\delta T_1\right\},\nonumber\\
   \Delta A_1^{(2)\alpha_s}&=&-\frac{2E}{m_B+m_{K^*}}\left\{C_F\alpha_s(\mu_b)\,\xi^{(2)}_{\perp}\,\left[\log\frac{m_b^2}{\mu_b^2}-L\right]\,+\,C_F\alpha_s(\mu_h)\delta T_1\right\},\nonumber\\
   \Delta A_2^{(2)\alpha_s}&=&-\frac{m_B}{m_B-m_{K^*}}\left\{C_F\alpha_s(\mu_b)\,\xi^{(2)}_{\perp}\,\left[\log\frac{m_b^2}{\mu_b^2}-L\right]\,-\,(\Delta-1)\xi^{(2)}_{\|}\,+\,C_F\alpha_s(\mu_h)\,\delta T_1\right\},\nonumber\\
   \Delta A_0^{(2)\alpha_s}&=&\Delta T_1^{(2)\alpha_s}\,=\,0,\nonumber\\
   \Delta T_2^{(2)\alpha_s}&=&C_F\alpha_s(\mu_h)\,\left(\delta T_2-\frac{2E}{m_B}\delta T_1\right),\nonumber\\
   \Delta T_3^{(2)\alpha_s}&=&\xi^{(2)}_{\|}\left\{\Delta\,\left(1+C_F\alpha_s(\mu_b)\left[\log\frac{m_b^2}{\mu_b^2}+2L\right]\right)-1\right\}\,+\,C_F\alpha_s(\mu_h)(\delta T_3 - \delta T_1).
\end{eqnarray}

\newpage

\section{SM predictions for other $B\to K^*\mu^+\mu^-$ observables}
\label{app2}

Here we collect the SM predictions for other observables not given in Section~\ref{sec:results}, as computed following the approach explored in this paper. Again, we collect for references the results in our preferred scheme 1, corresponding to defining the soft form
factors from $V$, $A_1$, $A_2$. 

\bigskip

\begin{table}[h]
\small
\centering
\begin{tabular}{@{}lrr@{}}
\hline\hline 
Observable & \begin{minipage}{6cm}\hspace{2cm} KMPW - scheme 1\end{minipage} & \begin{minipage}{6cm}\hspace{2cm} BZ - scheme 1 \end{minipage}  \\ 
 \hline 
$\av{P_3}_{[0.1,2]} $  & $-0.001_{-0.000}^{+0.000}{}_{-0.000}^{+0.000}{}_{-0.000}^{+0.000}{}_{-0.018}^{+0.017} $ & $-0.002_{-0.000}^{+0.000}{}_{-0.000}^{+0.000}{}_{-0.001}^{+0.000}{}_{-0.019}^{+0.017} $ \\ 
 \hline 
$\av{P_3}_{[2,4.3]} $  & $0.002_{-0.001}^{+0.001}{}_{-0.002}^{+0.004}{}_{-0.003}^{+0.003}{}_{-0.005}^{+0.005} $ & $-0.003_{-0.001}^{+0.000}{}_{-0.000}^{+0.000}{}_{-0.003}^{+0.003}{}_{-0.003}^{+0.003} $ \\ 
 \hline 
$\av{P_3}_{[4.3,8.68]} $  & $0.003_{-0.003}^{+0.000}{}_{-0.002}^{+0.005}{}_{-0.002}^{+0.003}{}_{-0.002}^{+0.003} $ & $-0.003_{-0.000}^{+0.001}{}_{-0.000}^{+0.000}{}_{-0.002}^{+0.002}{}_{-0.002}^{+0.003} $ \\ 
 \hline 
$\av{P_3}_{[1,6]} $  & $0.002_{-0.000}^{+0.001}{}_{-0.002}^{+0.003}{}_{-0.002}^{+0.003}{}_{-0.006}^{+0.005} $ & $-0.003_{-0.001}^{+0.000}{}_{-0.000}^{+0.000}{}_{-0.002}^{+0.002}{}_{-0.004}^{+0.004} $ \\ 
 \hline 
$\av{P_3}_{[1,2]} $  & $-0.001_{-0.000}^{+0.000}{}_{-0.000}^{+0.001}{}_{-0.002}^{+0.002}{}_{-0.017}^{+0.016} $ & $-0.003_{-0.001}^{+0.000}{}_{-0.000}^{+0.000}{}_{-0.002}^{+0.002}{}_{-0.018}^{+0.016} $ \\ 
 \hline 
$\av{P_3}_{[4.3,6]} $  & $0.003_{-0.001}^{+0.002}{}_{-0.002}^{+0.004}{}_{-0.002}^{+0.003}{}_{-0.001}^{+0.001} $ & $-0.002_{-0.001}^{+0.000}{}_{-0.000}^{+0.000}{}_{-0.002}^{+0.002}{}_{-0.001}^{+0.002} $ \\ 
 \hline 
$\av{P_3}_{[6,8]} $  & $0.002_{-0.004}^{+0.002}{}_{-0.002}^{+0.004}{}_{-0.002}^{+0.002}{}_{-0.002}^{+0.002} $ & $-0.002_{-0.001}^{+0.002}{}_{-0.000}^{+0.000}{}_{-0.001}^{+0.002}{}_{-0.002}^{+0.003} $ \\ 
 \hline 
$\av{P'_6}_{[0.1,2]} $  & $-0.071_{-0.030}^{+0.022}{}_{-0.013}^{+0.012}{}_{-0.006}^{+0.004}{}_{-0.015}^{+0.014} $ & $-0.074_{-0.031}^{+0.024}{}_{-0.001}^{+0.001}{}_{-0.006}^{+0.005}{}_{-0.016}^{+0.015} $ \\ 
 \hline 
$\av{P'_6}_{[2,4.3]} $  & $-0.084_{-0.036}^{+0.027}{}_{-0.020}^{+0.018}{}_{-0.002}^{+0.002}{}_{-0.008}^{+0.009} $ & $-0.084_{-0.035}^{+0.028}{}_{-0.001}^{+0.001}{}_{-0.002}^{+0.002}{}_{-0.008}^{+0.009} $ \\ 
 \hline 
$\av{P'_6}_{[4.3,8.68]} $  & $-0.067_{-0.020}^{+0.039}{}_{-0.022}^{+0.020}{}_{-0.003}^{+0.003}{}_{-0.013}^{+0.011} $ & $-0.063_{-0.016}^{+0.037}{}_{-0.001}^{+0.001}{}_{-0.003}^{+0.003}{}_{-0.014}^{+0.012} $ \\ 
 \hline 
$\av{P'_6}_{[1,6]} $  & $-0.076_{-0.036}^{+0.025}{}_{-0.019}^{+0.017}{}_{-0.002}^{+0.002}{}_{-0.007}^{+0.008} $ & $-0.075_{-0.033}^{+0.026}{}_{-0.001}^{+0.001}{}_{-0.002}^{+0.002}{}_{-0.007}^{+0.008} $ \\ 
 \hline 
$\av{P'_6}_{[1,2]} $  & $-0.089_{-0.036}^{+0.026}{}_{-0.016}^{+0.015}{}_{-0.007}^{+0.006}{}_{-0.016}^{+0.015} $ & $-0.093_{-0.037}^{+0.029}{}_{-0.001}^{+0.001}{}_{-0.007}^{+0.007}{}_{-0.017}^{+0.016} $ \\ 
 \hline 
$\av{P'_6}_{[4.3,6]} $  & $-0.061_{-0.039}^{+0.022}{}_{-0.020}^{+0.016}{}_{-0.003}^{+0.003}{}_{-0.007}^{+0.007} $ & $-0.058_{-0.036}^{+0.022}{}_{-0.001}^{+0.001}{}_{-0.003}^{+0.003}{}_{-0.008}^{+0.007} $ \\ 
 \hline 
$\av{P'_6}_{[6,8]} $  & $-0.059_{-0.045}^{+0.055}{}_{-0.021}^{+0.017}{}_{-0.003}^{+0.003}{}_{-0.014}^{+0.012} $ & $-0.056_{-0.045}^{+0.052}{}_{-0.001}^{+0.001}{}_{-0.003}^{+0.003}{}_{-0.015}^{+0.013} $ \\ 
 \hline 
$\av{P'_8}_{[0.1,2]} $  & $0.032_{-0.017}^{+0.027}{}_{-0.014}^{+0.015}{}_{-0.006}^{+0.007}{}_{-0.015}^{+0.014} $ & $0.034_{-0.017}^{+0.027}{}_{-0.001}^{+0.001}{}_{-0.006}^{+0.007}{}_{-0.016}^{+0.014} $ \\ 
 \hline 
$\av{P'_8}_{[2,4.3]} $  & $0.058_{-0.023}^{+0.036}{}_{-0.015}^{+0.019}{}_{-0.005}^{+0.005}{}_{-0.012}^{+0.010} $ & $0.057_{-0.023}^{+0.034}{}_{-0.001}^{+0.001}{}_{-0.004}^{+0.004}{}_{-0.012}^{+0.010} $ \\ 
 \hline 
$\av{P'_8}_{[4.3,8.68]} $  & $0.053_{-0.041}^{+0.018}{}_{-0.016}^{+0.020}{}_{-0.001}^{+0.001}{}_{-0.008}^{+0.006} $ & $0.049_{-0.037}^{+0.015}{}_{-0.001}^{+0.001}{}_{-0.002}^{+0.002}{}_{-0.008}^{+0.006} $ \\ 
 \hline 
$\av{P'_8}_{[1,6]} $  & $0.051_{-0.021}^{+0.035}{}_{-0.014}^{+0.018}{}_{-0.004}^{+0.004}{}_{-0.010}^{+0.009} $ & $0.050_{-0.021}^{+0.032}{}_{-0.001}^{+0.001}{}_{-0.003}^{+0.004}{}_{-0.010}^{+0.009} $ \\ 
 \hline 
$\av{P'_8}_{[1,2]} $  & $0.049_{-0.022}^{+0.035}{}_{-0.015}^{+0.017}{}_{-0.008}^{+0.009}{}_{-0.018}^{+0.015} $ & $0.052_{-0.023}^{+0.034}{}_{-0.001}^{+0.001}{}_{-0.008}^{+0.009}{}_{-0.018}^{+0.016} $ \\ 
 \hline 
$\av{P'_8}_{[4.3,6]} $  & $0.046_{-0.019}^{+0.037}{}_{-0.013}^{+0.017}{}_{-0.002}^{+0.002}{}_{-0.007}^{+0.006} $ & $0.043_{-0.018}^{+0.034}{}_{-0.000}^{+0.001}{}_{-0.002}^{+0.002}{}_{-0.007}^{+0.006} $ \\ 
 \hline 
$\av{P'_8}_{[6,8]} $  & $0.047_{-0.059}^{+0.038}{}_{-0.015}^{+0.019}{}_{-0.001}^{+0.001}{}_{-0.008}^{+0.007} $ & $0.044_{-0.053}^{+0.036}{}_{-0.001}^{+0.001}{}_{-0.001}^{+0.001}{}_{-0.008}^{+0.006} $ \\ 
 \hline 
$\av{A_\text{FB}}_{[0.1,2]} $  & $-0.131_{-0.001}^{+0.002}{}_{-0.058}^{+0.068}{}_{-0.004}^{+0.005}{}_{-0.000}^{+0.000} $ & $-0.123_{-0.002}^{+0.004}{}_{-0.006}^{+0.007}{}_{-0.005}^{+0.007}{}_{-0.000}^{+0.000} $ \\ 
 \hline 
$\av{A_\text{FB}}_{[2,4.3]} $  & $-0.080_{-0.013}^{+0.020}{}_{-0.085}^{+0.052}{}_{-0.033}^{+0.032}{}_{-0.004}^{+0.005} $ & $-0.047_{-0.013}^{+0.017}{}_{-0.003}^{+0.004}{}_{-0.033}^{+0.029}{}_{-0.003}^{+0.005} $ \\ 
 \hline 
$\av{A_\text{FB}}_{[4.3,8.68]} $  & $0.175_{-0.014}^{+0.024}{}_{-0.13}^{+0.173}{}_{-0.025}^{+0.022}{}_{-0.002}^{+0.002} $ & $0.204_{-0.012}^{+0.020}{}_{-0.014}^{+0.012}{}_{-0.024}^{+0.018}{}_{-0.002}^{+0.002} $ \\ 
 \hline 
$\av{A_\text{FB}}_{[1,6]} $  & $-0.042_{-0.012}^{+0.021}{}_{-0.027}^{+0.025}{}_{-0.031}^{+0.030}{}_{-0.003}^{+0.004} $ & $-0.013_{-0.010}^{+0.018}{}_{-0.000}^{+0.001}{}_{-0.031}^{+0.027}{}_{-0.003}^{+0.004} $ \\ 
 \hline 
$\av{A_\text{FB}}_{[1,2]} $  & $-0.199_{-0.006}^{+0.010}{}_{-0.21}^{+0.128}{}_{-0.019}^{+0.021}{}_{-0.002}^{+0.004} $ & $-0.174_{-0.009}^{+0.012}{}_{-0.014}^{+0.015}{}_{-0.018}^{+0.020}{}_{-0.003}^{+0.004} $ \\ 
 \hline 
$\av{A_\text{FB}}_{[4.3,6]} $  & $0.086_{-0.011}^{+0.025}{}_{-0.066}^{+0.121}{}_{-0.031}^{+0.028}{}_{-0.002}^{+0.002} $ & $0.118_{-0.010}^{+0.020}{}_{-0.009}^{+0.008}{}_{-0.030}^{+0.024}{}_{-0.002}^{+0.002} $ \\ 
 \hline 
$\av{A_\text{FB}}_{[6,8]} $  & $0.202_{-0.013}^{+0.028}{}_{-0.148}^{+0.184}{}_{-0.023}^{+0.021}{}_{-0.002}^{+0.002} $ & $0.231_{-0.011}^{+0.023}{}_{-0.015}^{+0.013}{}_{-0.022}^{+0.017}{}_{-0.002}^{+0.003} $ \\ 
 \hline 
\hline 
\end{tabular}
\caption{SM predictions for $P_3$, $P_6'$, $P_8'$, $A_\text{FB}$ in various bins. Same notation as Table~\ref{tab:res1}.}
\label{tab:res2}
\end{table}

\newpage

\begin{table}[h]
\small
\centering
\begin{tabular}{@{}lrr@{}}
\hline\hline 
Observable & \begin{minipage}{6cm}\hspace{2cm} KMPW - scheme 1\end{minipage} & \begin{minipage}{6cm}\hspace{2cm} BZ - scheme 1 \end{minipage}  \\ 
 \hline 
$\av{F_L}_{[0.1,2]} $  & $0.345_{-0.022}^{+0.028}{}_{-0.229}^{+0.278}{}_{-0.045}^{+0.050}{}_{-0.008}^{+0.010} $ & $0.400_{-0.026}^{+0.030}{}_{-0.024}^{+0.029}{}_{-0.045}^{+0.048}{}_{-0.009}^{+0.011} $ \\ 
 \hline 
$\av{F_L}_{[2,4.3]} $  & $0.763_{-0.009}^{+0.011}{}_{-0.294}^{+0.148}{}_{-0.021}^{+0.018}{}_{-0.003}^{+0.003} $ & $0.784_{-0.010}^{+0.011}{}_{-0.016}^{+0.018}{}_{-0.016}^{+0.016}{}_{-0.002}^{+0.002} $ \\ 
 \hline 
$\av{F_L}_{[4.3,8.68]} $  & $0.648_{-0.003}^{+0.006}{}_{-0.298}^{+0.244}{}_{-0.013}^{+0.012}{}_{-0.004}^{+0.004} $ & $0.638_{-0.006}^{+0.008}{}_{-0.021}^{+0.024}{}_{-0.015}^{+0.014}{}_{-0.004}^{+0.004} $ \\ 
 \hline 
$\av{F_L}_{[1,6]} $  & $0.717_{-0.010}^{+0.010}{}_{-0.305}^{+0.179}{}_{-0.022}^{+0.021}{}_{-0.003}^{+0.004} $ & $0.736_{-0.011}^{+0.011}{}_{-0.019}^{+0.021}{}_{-0.018}^{+0.019}{}_{-0.003}^{+0.003} $ \\ 
 \hline 
$\av{F_L}_{[1,2]} $  & $0.630_{-0.025}^{+0.030}{}_{-0.32}^{+0.203}{}_{-0.049}^{+0.048}{}_{-0.008}^{+0.010} $ & $0.688_{-0.026}^{+0.027}{}_{-0.021}^{+0.023}{}_{-0.044}^{+0.039}{}_{-0.007}^{+0.009} $ \\ 
 \hline 
$\av{F_L}_{[4.3,6]} $  & $0.710_{-0.004}^{+0.005}{}_{-0.302}^{+0.199}{}_{-0.013}^{+0.011}{}_{-0.002}^{+0.002} $ & $0.708_{-0.007}^{+0.008}{}_{-0.020}^{+0.022}{}_{-0.015}^{+0.013}{}_{-0.002}^{+0.002} $ \\ 
 \hline 
$\av{F_L}_{[6,8]} $  & $0.631_{-0.004}^{+0.007}{}_{-0.297}^{+0.257}{}_{-0.013}^{+0.013}{}_{-0.004}^{+0.004} $ & $0.617_{-0.006}^{+0.009}{}_{-0.021}^{+0.024}{}_{-0.016}^{+0.015}{}_{-0.004}^{+0.004} $ \\ 
 \hline 
$\av{S_3}_{[0.1,2]} $  & $0.005_{-0.000}^{+0.000}{}_{-0.003}^{+0.004}{}_{-0.002}^{+0.002}{}_{-0.010}^{+0.008} $ & $0.007_{-0.001}^{+0.001}{}_{-0.000}^{+0.000}{}_{-0.002}^{+0.002}{}_{-0.009}^{+0.007} $ \\ 
 \hline 
$\av{S_3}_{[2,4.3]} $  & $0.000_{-0.000}^{+0.000}{}_{-0.001}^{+0.000}{}_{-0.004}^{+0.004}{}_{-0.001}^{+0.001} $ & $-0.003_{-0.000}^{+0.000}{}_{-0.000}^{+0.000}{}_{-0.005}^{+0.004}{}_{-0.000}^{+0.000} $ \\ 
 \hline 
$\av{S_3}_{[4.3,8.68]} $  & $0.002_{-0.000}^{+0.000}{}_{-0.009}^{+0.005}{}_{-0.012}^{+0.011}{}_{-0.000}^{+0.000} $ & $-0.018_{-0.000}^{+0.000}{}_{-0.000}^{+0.001}{}_{-0.014}^{+0.013}{}_{-0.001}^{+0.000} $ \\ 
 \hline 
$\av{S_3}_{[1,6]} $  & $0.001_{-0.000}^{+0.000}{}_{-0.001}^{+0.001}{}_{-0.005}^{+0.004}{}_{-0.001}^{+0.001} $ & $-0.004_{-0.000}^{+0.000}{}_{-0.000}^{+0.000}{}_{-0.006}^{+0.005}{}_{-0.001}^{+0.001} $ \\ 
 \hline 
$\av{S_3}_{[1,2]} $  & $0.000_{-0.000}^{+0.000}{}_{-0.002}^{+0.005}{}_{-0.004}^{+0.003}{}_{-0.007}^{+0.005} $ & $0.004_{-0.000}^{+0.000}{}_{-0.000}^{+0.000}{}_{-0.002}^{+0.002}{}_{-0.005}^{+0.004} $ \\ 
 \hline 
$\av{S_3}_{[4.3,6]} $  & $0.002_{-0.000}^{+0.000}{}_{-0.006}^{+0.004}{}_{-0.009}^{+0.009}{}_{-0.000}^{+0.000} $ & $-0.011_{-0.000}^{+0.000}{}_{-0.000}^{+0.000}{}_{-0.011}^{+0.010}{}_{-0.000}^{+0.000} $ \\ 
 \hline 
$\av{S_3}_{[6,8]} $  & $0.002_{-0.000}^{+0.000}{}_{-0.009}^{+0.005}{}_{-0.012}^{+0.012}{}_{-0.000}^{+0.000} $ & $-0.020_{-0.000}^{+0.000}{}_{-0.001}^{+0.001}{}_{-0.015}^{+0.014}{}_{-0.001}^{+0.000} $ \\ 
 \hline 
$\av{S_4}_{[0.1,2]} $  & $-0.072_{-0.003}^{+0.003}{}_{-0.007}^{+0.028}{}_{-0.006}^{+0.006}{}_{-0.001}^{+0.001} $ & $-0.067_{-0.003}^{+0.004}{}_{-0.000}^{+0.000}{}_{-0.006}^{+0.008}{}_{-0.002}^{+0.002} $ \\ 
 \hline 
$\av{S_4}_{[2,4.3]} $  & $0.098_{-0.006}^{+0.007}{}_{-0.050}^{+0.040}{}_{-0.024}^{+0.025}{}_{-0.001}^{+0.001} $ & $0.123_{-0.006}^{+0.007}{}_{-0.005}^{+0.004}{}_{-0.024}^{+0.022}{}_{-0.001}^{+0.001} $ \\ 
 \hline 
$\av{S_4}_{[4.3,8.68]} $  & $0.212_{-0.002}^{+0.003}{}_{-0.086}^{+0.022}{}_{-0.015}^{+0.014}{}_{-0.001}^{+0.001} $ & $0.236_{-0.002}^{+0.002}{}_{-0.004}^{+0.003}{}_{-0.014}^{+0.012}{}_{-0.001}^{+0.001} $ \\ 
 \hline 
$\av{S_4}_{[1,6]} $  & $0.102_{-0.006}^{+0.007}{}_{-0.049}^{+0.036}{}_{-0.023}^{+0.024}{}_{-0.001}^{+0.001} $ & $0.125_{-0.006}^{+0.006}{}_{-0.004}^{+0.003}{}_{-0.023}^{+0.021}{}_{-0.001}^{+0.001} $ \\ 
 \hline 
$\av{S_4}_{[1,2]} $  & $-0.042_{-0.006}^{+0.008}{}_{-0.007}^{+0.013}{}_{-0.016}^{+0.021}{}_{-0.002}^{+0.002} $ & $-0.023_{-0.006}^{+0.009}{}_{-0.000}^{+0.000}{}_{-0.019}^{+0.020}{}_{-0.002}^{+0.002} $ \\ 
 \hline 
$\av{S_4}_{[4.3,6]} $  & $0.186_{-0.003}^{+0.004}{}_{-0.083}^{+0.034}{}_{-0.018}^{+0.017}{}_{-0.001}^{+0.001} $ & $0.212_{-0.003}^{+0.003}{}_{-0.005}^{+0.004}{}_{-0.017}^{+0.015}{}_{-0.001}^{+0.000} $ \\ 
 \hline 
$\av{S_4}_{[6,8]} $  & $0.221_{-0.002}^{+0.002}{}_{-0.088}^{+0.018}{}_{-0.013}^{+0.012}{}_{-0.001}^{+0.001} $ & $0.245_{-0.002}^{+0.001}{}_{-0.003}^{+0.002}{}_{-0.012}^{+0.011}{}_{-0.001}^{+0.001} $ \\ 
 \hline 
$\av{S_5}_{[0.1,2]} $  & $0.207_{-0.007}^{+0.004}{}_{-0.061}^{+0.008}{}_{-0.016}^{+0.012}{}_{-0.005}^{+0.004} $ & $0.211_{-0.009}^{+0.006}{}_{-0.001}^{+0.000}{}_{-0.018}^{+0.013}{}_{-0.005}^{+0.005} $ \\ 
 \hline 
$\av{S_5}_{[2,4.3]} $  & $-0.167_{-0.025}^{+0.018}{}_{-0.038}^{+0.068}{}_{-0.035}^{+0.040}{}_{-0.007}^{+0.006} $ & $-0.172_{-0.023}^{+0.016}{}_{-0.004}^{+0.006}{}_{-0.032}^{+0.034}{}_{-0.006}^{+0.005} $ \\ 
 \hline 
$\av{S_5}_{[4.3,8.68]} $  & $-0.424_{-0.019}^{+0.011}{}_{-0.021}^{+0.157}{}_{-0.015}^{+0.018}{}_{-0.002}^{+0.002} $ & $-0.406_{-0.016}^{+0.010}{}_{-0.005}^{+0.007}{}_{-0.018}^{+0.019}{}_{-0.002}^{+0.002} $ \\ 
 \hline 
$\av{S_5}_{[1,6]} $  & $-0.178_{-0.027}^{+0.016}{}_{-0.040}^{+0.069}{}_{-0.033}^{+0.038}{}_{-0.006}^{+0.005} $ & $-0.177_{-0.024}^{+0.015}{}_{-0.004}^{+0.005}{}_{-0.031}^{+0.032}{}_{-0.006}^{+0.005} $ \\ 
 \hline 
$\av{S_5}_{[1,2]} $  & $0.149_{-0.023}^{+0.016}{}_{-0.045}^{+0.011}{}_{-0.041}^{+0.039}{}_{-0.008}^{+0.007} $ & $0.135_{-0.024}^{+0.018}{}_{-0.004}^{+0.003}{}_{-0.041}^{+0.038}{}_{-0.008}^{+0.007} $ \\ 
 \hline 
$\av{S_5}_{[4.3,6]} $  & $-0.369_{-0.024}^{+0.011}{}_{-0.038}^{+0.148}{}_{-0.020}^{+0.023}{}_{-0.002}^{+0.002} $ & $-0.359_{-0.020}^{+0.010}{}_{-0.007}^{+0.009}{}_{-0.021}^{+0.022}{}_{-0.002}^{+0.002} $ \\ 
 \hline 
$\av{S_5}_{[6,8]} $  & $-0.445_{-0.021}^{+0.011}{}_{-0.019}^{+0.161}{}_{-0.015}^{+0.016}{}_{-0.001}^{+0.002} $ & $-0.424_{-0.018}^{+0.010}{}_{-0.004}^{+0.006}{}_{-0.018}^{+0.019}{}_{-0.001}^{+0.002} $ \\ 
 \hline 
$\av{S_{6s}}_{[0.1,2]} $  & $0.174_{-0.003}^{+0.002}{}_{-0.09}^{+0.078}{}_{-0.007}^{+0.006}{}_{-0.000}^{+0.000} $ & $0.163_{-0.005}^{+0.003}{}_{-0.009}^{+0.008}{}_{-0.009}^{+0.007}{}_{-0.001}^{+0.001} $ \\ 
 \hline 
$\av{S_{6s}}_{[2,4.3]} $  & $0.105_{-0.026}^{+0.018}{}_{-0.069}^{+0.113}{}_{-0.043}^{+0.044}{}_{-0.007}^{+0.005} $ & $0.062_{-0.023}^{+0.017}{}_{-0.005}^{+0.005}{}_{-0.039}^{+0.044}{}_{-0.006}^{+0.005} $ \\ 
 \hline 
$\av{S_{6s}}_{[4.3,8.68]} $  & $-0.235_{-0.032}^{+0.018}{}_{-0.231}^{+0.173}{}_{-0.030}^{+0.034}{}_{-0.003}^{+0.003} $ & $-0.273_{-0.027}^{+0.016}{}_{-0.016}^{+0.018}{}_{-0.025}^{+0.033}{}_{-0.003}^{+0.003} $ \\ 
 \hline 
$\av{S_{6s}}_{[1,6]} $  & $0.055_{-0.029}^{+0.016}{}_{-0.033}^{+0.036}{}_{-0.041}^{+0.042}{}_{-0.006}^{+0.004} $ & $0.017_{-0.025}^{+0.013}{}_{-0.001}^{+0.001}{}_{-0.037}^{+0.042}{}_{-0.005}^{+0.004} $ \\ 
 \hline 
$\av{S_{6s}}_{[1,2]} $  & $0.265_{-0.014}^{+0.008}{}_{-0.171}^{+0.28}{}_{-0.028}^{+0.025}{}_{-0.005}^{+0.003} $ & $0.231_{-0.017}^{+0.013}{}_{-0.020}^{+0.018}{}_{-0.027}^{+0.024}{}_{-0.005}^{+0.004} $ \\ 
 \hline 
$\av{S_{6s}}_{[4.3,6]} $  & $-0.116_{-0.033}^{+0.015}{}_{-0.161}^{+0.089}{}_{-0.038}^{+0.042}{}_{-0.003}^{+0.003} $ & $-0.159_{-0.026}^{+0.013}{}_{-0.011}^{+0.012}{}_{-0.033}^{+0.041}{}_{-0.003}^{+0.002} $ \\ 
 \hline 
$\av{S_{6s}}_{[6,8]} $  & $-0.270_{-0.037}^{+0.017}{}_{-0.246}^{+0.198}{}_{-0.028}^{+0.031}{}_{-0.003}^{+0.003} $ & $-0.309_{-0.030}^{+0.015}{}_{-0.017}^{+0.020}{}_{-0.022}^{+0.030}{}_{-0.004}^{+0.003} $ \\ 
 \hline 
\hline 
\end{tabular}
\caption{SM predictions for $F_L$, $S_3$, $S_4$, $S_5$, $S_{6s}$ in various bins. Same notation as Table~\ref{tab:res1}.}
\label{tab:res3}
\end{table}

\end{document}